\documentclass{article}

\PassOptionsToPackage{numbers,sort&compress}{natbib}
 \usepackage[preprint]{neurips_2026}

\usepackage[utf8]{inputenc} % allow utf-8 input
\usepackage[T1]{fontenc}    % use 8-bit T1 fonts
\usepackage{hyperref}       % hyperlinks
\usepackage{url}            % simple URL typesetting
\usepackage{booktabs}       % professional-quality tables
\usepackage{amsfonts}       % blackboard math symbols
\usepackage{nicefrac}       % compact symbols for 1/2, etc.
\usepackage{microtype}      % microtypography
\usepackage{xcolor}         % colors

\usepackage{amsmath,amssymb}
\usepackage{graphicx}
\usepackage{booktabs}
\usepackage{arydshln}  % dashed lines in tables
\usepackage{multirow}
\usepackage{xcolor}
\usepackage{hyperref}
\usepackage{natbib}
\usepackage{geometry}
\DeclareSymbolFont{extraup}{U}{zavm}{m}{n}
\DeclareMathSymbol{\varheart}{\mathord}{extraup}{86} % 86 = "56(hex)

\title{SURGE: An Event-Centric Social Media Sentiment Time Series Benchmark with Interaction Structure}

\author{%
  Chen Su$^{{\spadesuit}}$, \hspace{0.1cm}
  Pengsen Cheng$^{\clubsuit}$, \hspace{0.1cm}
    Yuanhe Tian$^{\varheart}$, \hspace{0.1cm}
    Yan Song$^{{\spadesuit}*}$
    \\
    $^{\spadesuit}$University of Science and Technology of China \\
    $^{\clubsuit}$Sichuan University \hspace{0.1cm}
    $^{\varheart}$Zhongguancun Academy
    \\
    $^{\spadesuit}$\texttt{suchen4565@mail.ustc.edu.cn} \hspace{0.1cm}
    $^{\clubsuit}$\texttt{chengpengsen@scu.edu.cn} \\
    $^{\varheart}$\texttt{yhtian94@gmail.com} \hspace{0.1cm}
    $^{\spadesuit}$\texttt{clksong@gmail.com}
}
\begin{document}

\renewcommand{\thefootnote}{\fnsymbol{footnote}}
\footnotetext[1]{Corresponding author.}

\renewcommand{\thefootnote}{\arabic{footnote}}

\maketitle

% Abstract
\begin{abstract}

Public events on social media generate large volumes of discussion whose collective dynamics carry direct value for opinion forecasting and crisis response. Capturing how these dynamics evolve across an event's lifecycle requires organizing fragmented posts into event-level time series. Existing datasets cover only a small number of events within a single category, and typically discard the interaction structure between posts when constructing time series, which restricts both transfer across event types and controlled study of how interactions shape the resulting collective dynamics. We present SURGE, a multi-event social media benchmark that pairs event-level time series with aligned text and interaction structure linking posts within an event. SURGE is built through an automated pipeline that produces calendar-aligned time series at three temporal granularities, covering 67 events and more than 800K posts across five event categories. Each time bin is paired with flat and structured textual views derived from the same selected posts, enabling controlled evaluation of whether social interaction structure affects forecasting behavior. On top of SURGE we define benchmark protocols for numerical-only forecasting, text-augmented forecasting, high-interaction evaluation, and leave-one-category-out generalization. Experiments with representative time-series and multimodal forecasting models reveal three properties of the benchmark: a strong local-persistence regime in which naive baselines remain hard to beat under absolute error, limited transfer of existing text-augmented forecasters to event-driven social-media data, and increased difficulty on reply-dense periods that aggregate metrics tend to obscure. We further include a lightweight structure-aware probe as a reference implementation, illustrating how SURGE can support interaction-aware forecasting research.\footnote{The dataset and code are released at \url{https://github.com/synlp/SURGE}.}

\end{abstract}

% Main sections
\section{Introduction}

Public events such as natural disasters, policy controversies, social movements, and technology releases generate massive volumes of discussion on social media \cite{mao2012correlating, bollen2011twitter, cinelli2020infodemic, conover2011polarization, lehmann2012dynamical, pinto2019dynamical, noellenuemann1974spiral}.
Within this discussion, the moments when conversation networks rapidly densify and collective sentiment shifts most abruptly are usually the more consequential ones, since crisis response, policy assessment, and opinion monitoring all depend on anticipating such shifts before they fully unfold rather than describing them in hindsight \cite{nguyen2012predicting, naskar2020emotiondynamics, zhang2024decoding}.
Anticipation under this requirement calls for predictive modeling over the temporal evolution of public discussion.

Predictive modeling of this kind imposes three demands on the underlying data.
Public discussion has to be organized at the event level, since the dynamics being predicted unfold along an event's lifecycle rather than along an unbounded global stream.
Calendar-aligned time series provide the predictive interface that this lifecycle-level data must adopt, admitting comparable evaluation across models.
Beyond this temporal form, the reply and repost topology among posts has to accompany the time series, because in event-driven social media the temporal dynamics are tightly coupled with the interaction structure that connects posts within an event \cite{galesic2021opinion, demszky2019tracking, degroot1974consensus, friedkin1999fjmodel, bernardo2024boundedconfidence, dandekar2013biasedassim, proskurnikov2017tutorial1, xu2025sentimentdiffusion, zhang2024ntom, hu2024resemo,tian2024emotion}, rather than driven by an external textual signal independent of the audience.

Existing resources only partially satisfy these demands.
Social media event datasets often capture event boundaries and retain reply structure \cite{hu2020weibocov, zubiaga2016pheme, chen2024crosscultural, shu2018fakenewsnet, ma2018rumortree, bian2020rumorgcn, li2025rhythm}, while their constructed time series either anchor the signal to an exogenous driver such as weather rather than to the discussion itself \cite{meunier2025crisisTS}, or organize the series by individual user rather than by event \cite{cai2023depression}, so cross-event transferability of how interaction structure shapes discussion dynamics cannot be tested on any of them.
Multimodal time series forecasting datasets supply numerical signals paired with aligned text in a forecasting setup, but their text comes from news reports, financial filings, or government bulletins, where text functions as an exogenous side channel without an addressee, and reply or repost topology of the kind found in social media has no analogue in this regime \cite{anonymous2024mmtsflib, chang2025timeimm, ni2026streasoner, wang2024newstoforecast, chen2025mtbench, dong2024fnspid, xu2018stocknet,su2025text}.
No existing benchmark therefore brings event-organized, discussion-derived time series and the reply structure of the underlying discussion together, leaving the predictive value of interaction structure for collective opinion dynamics untested.

To address this gap, we present SURGE, an event-centric social media benchmark that pairs event-level temporal signals with the reply and repost structure of the underlying discussion across multiple event categories, designed for forecasting rather than description.
Our contributions are threefold.
First, we introduce SURGE, an event-centric social media benchmark that organizes 67 public events into calendar-aligned time series at three temporal granularities, including 1-day, 12-hour, and 6-hour resolutions, paired with aligned textual views and reply and repost interaction structure, constructed through an automated pipeline that unifies heterogeneous data from Twitter, Reddit, and Threads and annotates sentiment with large language models (LLMs).
Second, we define a set of evaluation protocols that cover within-event forecasting, text-augmented forecasting under matched configurations without text, with flat text, and with structured text, high-interaction-period evaluation, and leave-one-category-out generalization across the five event categories of natural disasters, political events, social movements, technology releases, and sports and entertainment events.
Third, we provide an extensive benchmark study with representative numerical and text-augmented forecasters, showing that SURGE exposes a strong persistence regime, limited robustness of existing multimodal time series forecasting (TSF) models under social-media text, and substantial difficulty shifts across interaction density and event category. We also release a lightweight structure-aware probe as a reference implementation for future interaction-aware models, rather than as a method-level contribution.

\vspace{-0.2cm}
\section{Related Work}
\vspace{-0.1cm}
Existing datasets most relevant to our work fall into two complementary lineages, one organized around social media event dynamics and the other around general text-augmented time series forecasting.
Within the first lineage, Weibo-COV provides a large collection of COVID-19-related Weibo posts with user-level and content-level metadata, yet it is restricted to a single event and releases only raw posts rather than pre-built event-level time series \cite{hu2020weibocov}.
CrisisTS couples meteorological time series with crisis-related tweets across multiple events, but its time series is an exogenous meteorological signal rather than a sentiment- or volume-based signal derived from the tweets itself, and it does not preserve interaction structure~\cite{meunier2025crisisTS}.
CovidSEE\&CovidSEC strengthen particular analytic dimensions in cross-cultural crisis response comparison, but cover only 2 events and are released as raw tweets without pre-built time series \cite{chen2024crosscultural}.
PHEME preserves conversation-thread structure around 9 breaking-news events, but its target is rumour verification rather than sentiment forecasting, and it releases no time series \cite{zubiaga2016pheme}.
VISTA preserves a three-level reply hierarchy across 159 Weibo trending topics with per-comment 11-class sentiment annotations, but releases per-comment arrival timestamps for Hawkes-process modeling rather than calendar-aligned aggregated time series, framing the task as point-process arrival prediction rather than forecasting over per-step aggregated signals \cite{li2025rhythm}.
SWDD is organized by user rather than by event and supports per-user depressive-symptom time series, but its organizational unit is the individual rather than the public event, and no interaction structure is retained~\cite{cai2023depression}.
The second lineage pairs numerical time series with aligned text. Time-MMD and Time-IMM organize data by domain rather than event, with text drawn from reports or news feeds rather than social discussion, and preserve no interaction structure between textual units~\cite{anonymous2024mmtsflib, chang2025timeimm}.
STReasoner introduces graph-defined spatial structure alongside aligned text, but its graphs encode physical adjacency over infrastructural systems such as traffic, power, and river networks, and both the time series and the descriptions are synthesized from network stochastic differential equations (SDEs) and template-based agents, rather than reflecting real social discussions linked by reply or repost interactions~\cite{ni2026streasoner}.
Taken together across both lineages, no existing dataset simultaneously provides event-level multi-event coverage, pre-built time series, bin-aligned textual views, and explicit social interaction structure.
In contrast, SURGE provides a pre-built time series spanning 67 events, along with an aligned text preserving interactive structure.
\begin{table*}
\centering
\caption{Comparison of SURGE with representative social media event datasets and text-multimodal time series forecasting datasets. ``Pre-built Time Series'' indicates whether time series values are released. ``Aligned Text'' indicates whether text is paired with the time series. ``Interaction Structure'' indicates whether reply/repost relations between posts are preserved.}
\label{tab:dataset_comparison}
\resizebox{\textwidth}{!}{%
\begin{tabular}{llcccc}
\toprule
\textbf{Dataset} & \textbf{Organization} & \textbf{Post Scale} & \textbf{Pre-built Time Series} & \textbf{Aligned Text} & \textbf{Interaction Structure} \\
\midrule
\multicolumn{6}{l}{\textit{Social media event datasets}} \\
Weibo-COV \cite{hu2020weibocov}            & 1 event           & 40M     & \texttimes & \texttimes & \texttimes \\
CrisisTS \cite{meunier2025crisisTS}        & 29 events  & 22K     & \checkmark & \checkmark & \texttimes \\
PHEME \cite{zubiaga2016pheme}              & 9 news   & 4.8K    & \texttimes & \texttimes & \checkmark \\
VISTA \cite{li2025rhythm}                  & 159 topics & 404K   & \texttimes & \texttimes & \checkmark \\
CovidSEE\&CovidSEC\cite{chen2024crosscultural} & 2 events        & 97K      & \texttimes & \texttimes & \texttimes \\
SWDD \cite{cai2023depression}              & 23K users         & 4.85M  & \checkmark & \texttimes & \texttimes \\
\midrule
\multicolumn{6}{l}{\textit{Text-multimodal time series forecasting datasets}} \\
Time-MMD \cite{anonymous2024mmtsflib}      & 9 domains         & --        & \checkmark & \checkmark & \texttimes \\
Time-IMM \cite{chang2025timeimm}           & 9 sub-datasets    & --        & \checkmark & \checkmark & \texttimes \\
STReasoner \cite{ni2026streasoner}         & 10 domains        & --        & \checkmark & \checkmark & \texttimes \\
\midrule
\textbf{SURGE (ours)}               & \textbf{67 events} & \textbf{817K} & \checkmark & \checkmark & \checkmark \\
\bottomrule
\end{tabular}%
\vspace{-0.2cm}
}
\end{table*}
\vspace{-0.2cm}
\section{The SURGE Benchmark}
\label{sec:dataset}
\vspace{-0.2cm}
\subsection{Data Source and Processing}
\label{sec:data_source_processing}
\label{sec:data_collection}
\label{sec:collection_filtering}

We target five categories of public events, specifically natural disasters, political events, social movements, technology releases, and sports and entertainment events.
On data scale, SURGE strikes a balance between two extremes in existing event-level social media datasets.
At one extreme are single-event very-large-scale corpora such as Weibo-COV \cite{hu2020weibocov} with 40M posts focused on COVID-19, and at the other are multi-event small-scale corpora such as CrisisTS \cite{meunier2025crisisTS} with 22K posts across 29 crisis events.
Collection spans three platforms, namely Twitter, Reddit, and Threads, to broaden the coverage of topic types and user populations.
Raw collection under this coverage plan yielded 93 candidate events and 1{,}256{,}816 posts.

\vspace{-0.2cm}
\paragraph{Processing.}
Posts in the raw collection are first deduplicated by identifier and filtered to remove records that are excessively short, dominated by emojis or uniform resource locators (URLs), or non-English.
This step removes low-signal content that does not contribute meaningfully to event-level sentiment or volume aggregation.
The complete rule set, including event-level retention thresholds, is documented in Appendix~\ref{sec:appendix_schema_filtering}.
Public events typically contain long sparse periods before and after the main discussion, in which the few scattered posts reflect ambient chatter loosely tied to the event itself rather than substantive engagement with it.
We therefore identify each event's active period and retain only the posts whose timestamps fall within it.
After both filtering stages, the curated corpus contains 67 events and 817{,}442 posts distributed across five categories, namely 17 political events, 12 natural disasters, 12 social movements, 12 technology events, and 14 sports and entertainment events.
We defer the active-period detection procedure to Appendix~\ref{sec:appendix_ts_construction}.
Each retained post is then assigned one of three sentiment labels, namely positive, neutral, or negative, mapped to numerical scores in $\{+1, 0, -1\}$.
This three-way scheme provides a practical balance between expressiveness and annotation stability \cite{nakov2016semeval,chen2020joint,qin2021improving}.
We obtain these labels using Qwen3-32B\footnote{\url{https://huggingface.co/Qwen/Qwen3-32B}} \cite{yang2025qwen3} in a zero-shot manner, and the resulting labels serve as the base signal for time series construction.
To assess label reliability and detect category- or class-specific drift, we conduct a human verification study on $3{,}000$ posts stratified jointly across the five event categories and the three sentiment classes.
Two annotators independently relabel each post under a shared codebook without access to the LLM-assigned labels, achieving a Cohen's $\kappa$ \cite{cohen1960kappa} of $0.74$ between annotators, and the LLM labels match the human consensus on $86.5\%$ of posts overall, with the per-stratum systematic bias bounded by $|\mu| \leq 0.05$.
The full protocol, including the prompt template, codebook definitions, and per-class results, appears in Appendix~\ref{sec:appendix_data_preprocessing}.

\begin{figure*}[t]
\centering
\includegraphics[width=\textwidth]{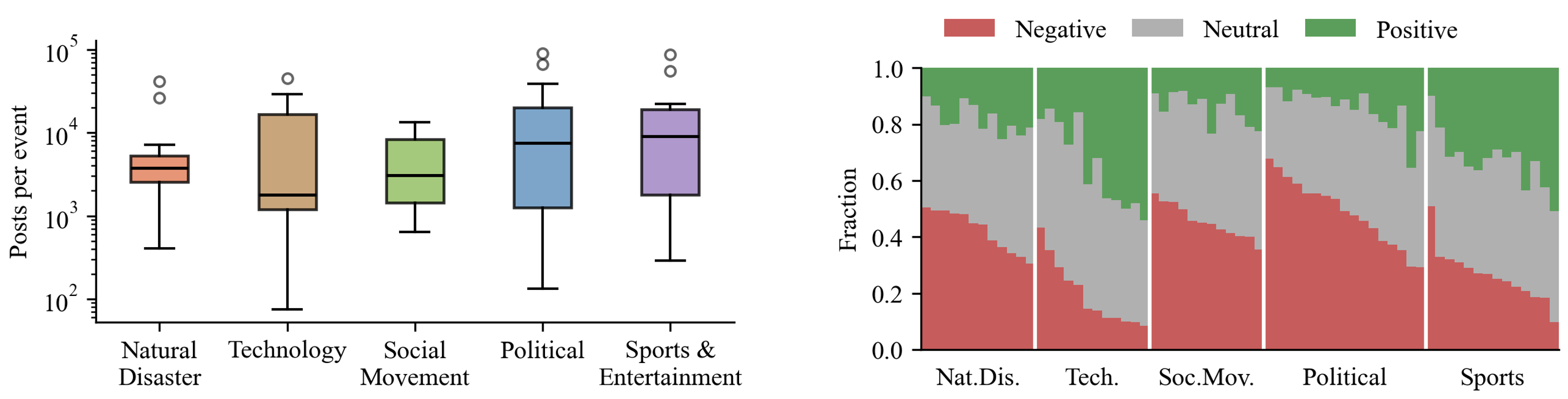}
\caption{Statistical overview of SURGE on the text side. The left panel shows the distribution of post volume across event categories. The right panel visualizes event-level sentiment distribution grouped by category.}
\label{fig:text_stats}
\vspace{-0.2cm}
\end{figure*}

\vspace{-0.2cm}
\paragraph{Distributional Characteristics.}
Spanning multiple events across the five categories listed above, the curated corpus already carries substantial event-level variability in both post volume and sentiment composition.
As shown in the left panel of Figure~\ref{fig:text_stats}, different event categories exhibit markedly different degrees of within-category variance in post volume.
Political and sports or entertainment events display the widest spread, with interquartile ranges spanning nearly an order of magnitude and individual events reaching volumes far above the category median, whereas natural disaster and technology events cluster more tightly.
A parallel pattern emerges in sentiment distribution, as depicted in the right panel of Figure~\ref{fig:text_stats}.
Natural disasters and political events tend toward more negative sentiment overall, whereas technology and sports or entertainment events more often exhibit higher neutral or positive fractions.
Crucially, this variation is not limited to cross-category contrasts. Events within the same category still display noticeably different sentiment mixtures.

\vspace{-0.2cm}
\subsection{Benchmark Design}
\label{sec:benchmark_design}

The curated corpus established in Section~\ref{sec:data_source_processing} provides per-post content and labels at the event level, but characterizing how collective discussion evolves across an event's lifecycle, and anticipating shifts in that evolution before they fully unfold, requires aggregating these posts into calendar-aligned signals on which forecasting models can operate.
We therefore construct a second data layer on top of the curated corpus, in which each event is represented by per-bin numerical signals together with bin-aligned textual views and the interaction structure linking posts within each bin.

\begin{figure}[t]
    \centering
    \includegraphics[width=\linewidth]{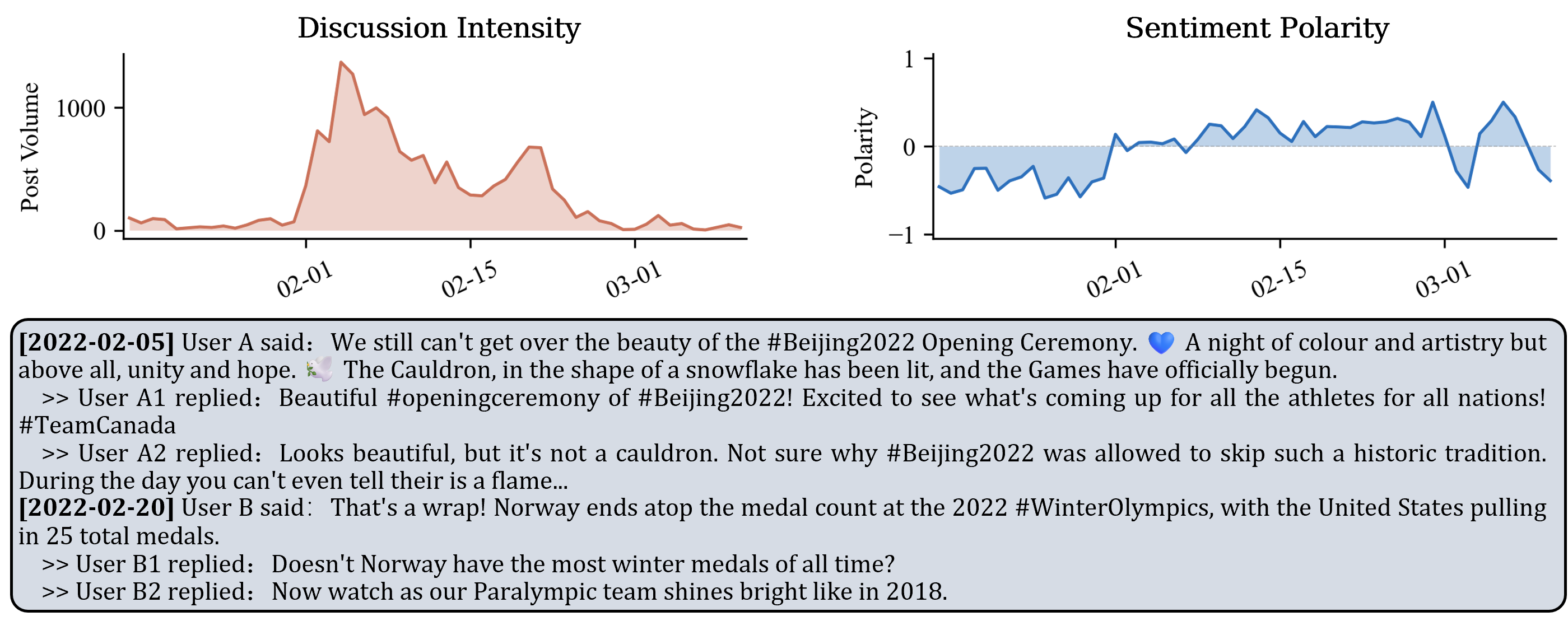}
    \caption{An illustrative SURGE record on Beijing Winter Olympics 2022. The top-left panel shows the Discussion Intensity (post volume), and the top-right panel shows the Sentiment Polarity. The bottom panel displays the structured text, in which replies are nested under their parent posts, preserving the reply-chain hierarchy.}
    \label{fig:data_case}
    \vspace{-0.2cm}
\end{figure}

\vspace{-0.2cm}
\paragraph{Time Series Construction.}
We adopt fixed-interval, calendar-aligned temporal bins and construct event-level time series at three temporal resolutions, namely 1 day, 12 hours, and 6 hours. These resolutions together capture event dynamics ranging from longer-term trends to finer-grained fluctuations. Each event is required to contain enough bins at a given granularity for the active period to yield at least one forecasting window assigned to the test split under chronological train/validation/test partitioning. Because this minimum-bin requirement translates into shorter calendar-duration thresholds at finer granularities, short-burst events that fail this requirement at the 1-day resolution still satisfy it at finer resolutions, yielding a per-granularity coverage of 55 events at 1-day, 64 at 12-hour, and 67 at 6-hour rather than a single fixed event set across granularities.
Going finer than 6 hours thins per-bin post counts and destabilizes aggregated signals, so we cap the resolution there.
Let $\mathcal{P}_t$ denote the set of posts assigned to time bin $t$, and let $s_p \in \{-1,\,0,\,+1\}$ denote the sentiment score of post $p$, corresponding to negative, neutral, and positive sentiment, respectively. We derive two target variables from each bin, namely the \emph{Discussion Intensity} ($c_t$) and the \emph{Sentiment Polarity} ($\bar{s}_t$) shown in the top panels of Figure~\ref{fig:data_case}.
\begin{equation}
c_t \;=\; |\mathcal{P}_t|, \qquad
\bar{s}_t \;=\; \frac{1}{c_t}\sum_{p \in \mathcal{P}_t} s_p,
\label{eq:targets}
\end{equation}
The two targets are normalized by per-event z-score.
Appendix~\ref{sec:appendix_ts_construction} describes the binning rules, active-period detection, and per-event normalization in full.

\begin{figure*}[t]
\centering
\includegraphics[width=\textwidth]{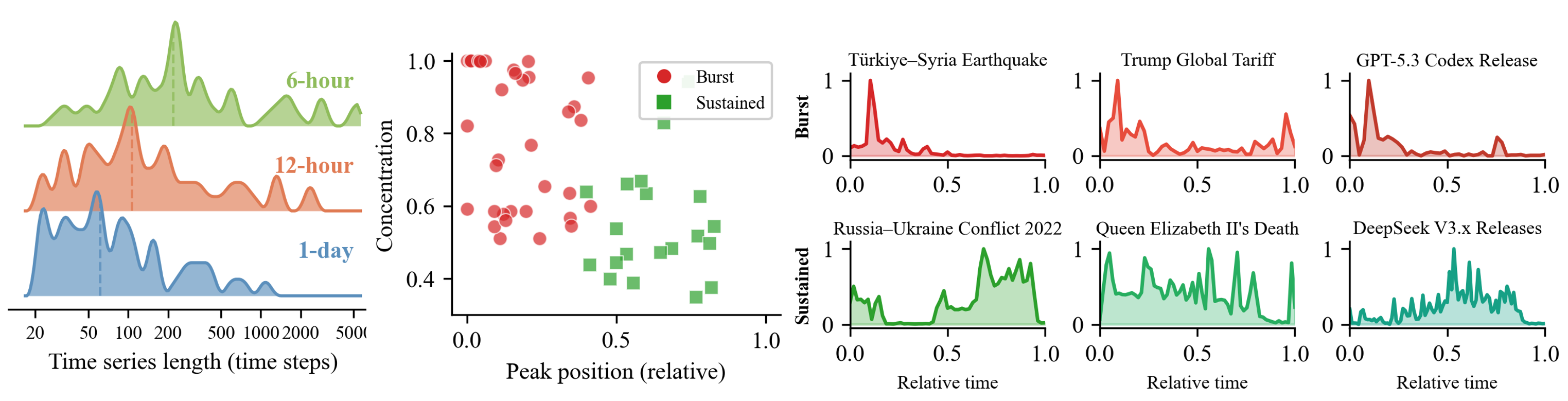}
\caption{Statistical overview of SURGE on the time-series side. The left panel summarizes time series length distributions across the three temporal resolutions. The middle panel maps events in a low-dimensional morphology space defined by relative peak position and activity concentration, revealing two dominant dynamical regimes. The right panel shows representative post-volume trajectories for the two regimes where the volume is normalized by the event peak.}
\label{fig:dataset_stats}
\end{figure*}

\vspace{-0.2cm}
\paragraph{Interaction Structure and Textual Views.}
Social media discussions are not isolated posts, but conversational structures formed by reply and repost relations. SURGE preserves these relations explicitly through bin-aligned interaction edges and through two paired textual views for each time bin, both constructed from the same underlying content. The structured view, illustrated in the bottom panel of Figure~\ref{fig:data_case}, organizes the selected posts into reply chains, nesting each reply under its parent post with explicit role markers, whereas the flat view linearizes the same posts in chronological order without role markers.
Following prior conversation-thread datasets such as PHEME, which organize quotes and replies under a unified thread structure, we construct the structured view by treating reposts as reply-like interactions. Specifically, replies and reposts are both organized into reply chains, since both capture user responses to an originating post and play similar roles in information diffusion.\footnote{For brevity, we use ``reply'' and ``reply chain'' throughout the remainder of the paper to refer to both reply and repost interactions, unless explicitly stated otherwise.}
The post-selection rule and text formatting are spelled out in Appendix~\ref{sec:appendix_ts_construction}.

\vspace{-0.2cm}
\paragraph{Task Formalization.}
On these per-bin targets and paired textual views, we formalize the benchmark as a multimodal time series forecasting task.
Given an event $e$ with historical numerical series $\mathbf{x}_{1:L}^{(e)} = (x_1, x_2, \ldots, x_L)$ over the past $L$ temporal bins and aligned bin-level textual observations $\mathbf{t}_{1:L}^{(e)} = (t_1, t_2, \ldots, t_L)$ taking either of the two views above, the model predicts the next $H$ values $\hat{\mathbf{x}}_{L+1:L+H}^{(e)} = f\bigl(\mathbf{x}_{1:L}^{(e)},\,\mathbf{t}_{1:L}^{(e)}\bigr)$.
A model that ignores $\mathbf{t}^{(e)}$ entirely reduces the task to standard numerical forecasting, while a model that consumes the textual view takes either the flat or the structured form defined above.
% 除事件内预测外，benchmark 还包含跨类别泛化协议。
Beyond within-event forecasting, the benchmark also includes a cross-category generalization protocol that holds out all events from one category for testing and uses the remaining categories for training, examining whether learned mappings transfer across event types.

\vspace{-0.2cm}
\paragraph{Forecasting Challenges.}
Each series in SURGE is anchored to its event's lifecycle, with length tracking how long that event sustains public attention.
Lengths therefore span from days to months and form a strongly right-skewed distribution (left panel of Figure~\ref{fig:dataset_stats}), shifting further right under finer granularities, in contrast with standard time series benchmarks whose fixed collection windows yield roughly comparable lengths across instances.
Trajectory shape similarly reflects how public attention concentrates around the event.
Morphology clustering on event-level post-volume trajectories resolves two qualitatively distinct regimes in the middle and right panels of Figure~\ref{fig:dataset_stats}.
The first is burst-like, with discussion peaking early and decaying rapidly.
The second corresponds to sustained dynamics, where attention spreads over a longer horizon with frequent reactivation.
Event categories correlate only loosely with regime, natural disasters and technology releases leaning toward burst and social movements toward sustained.
Both length heterogeneity and trajectory irregularity trace back to a single source, namely that the series is driven by how public attention rises and falls around an individual event rather than by the regular processes underlying conventional time series data.
Past values therefore form only part of the signal needed for forecasting, and the multimodal input space formalized above admits per-event textual content and the interaction structure connecting posts within the event as complementary signal grounded in the event itself.

\begin{figure*}[t]
\centering
\includegraphics[width=\linewidth]{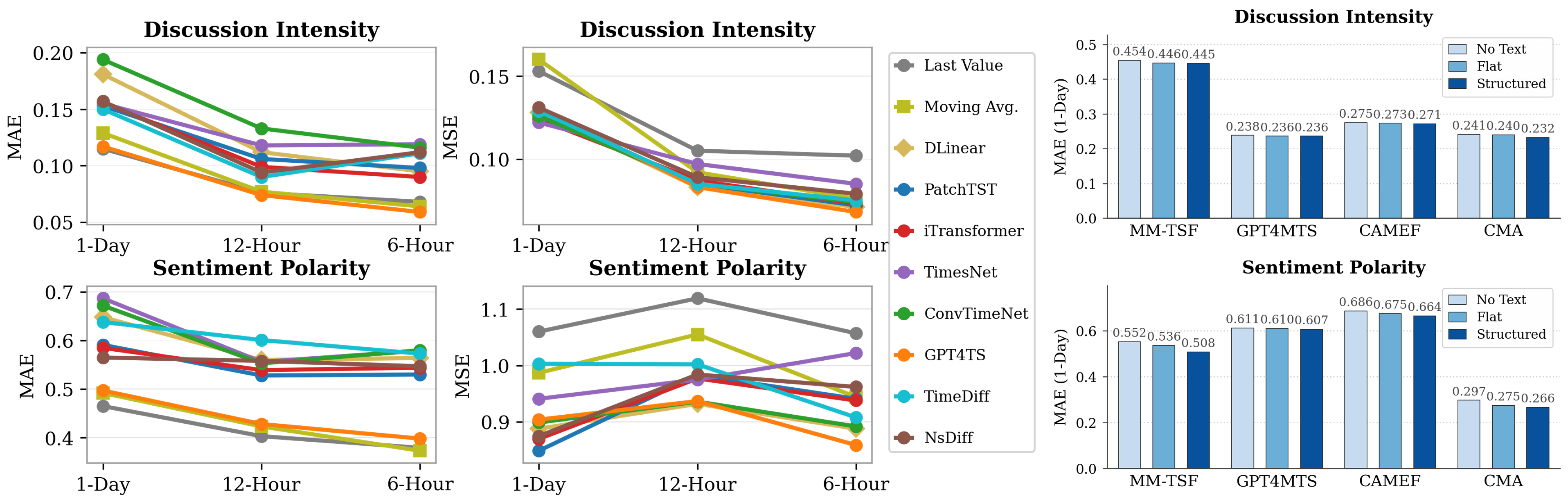}
\caption{Numerical-only forecasting and text-augmented forecasting performance. \textbf{Left:} numerical forecasting MAE and MSE on Discussion Intensity and Sentiment Polarity at three temporal granularities, namely 1-day (1D), 12-hour (12H), and 6-hour (6H), with one curve per model. \textbf{Right:} text-augmented MAE for the four multimodal models, namely MM-TSF, GPT4MTS, CAMEF, and the cross-modal attention (CMA) probe, under three text configurations (No Text, Flat, Structured) on Discussion Intensity and Sentiment Polarity at 1-day.}
\label{fig:num_text_results}
\end{figure*}

\begin{figure*}[t]
\centering
\includegraphics[width=\linewidth]{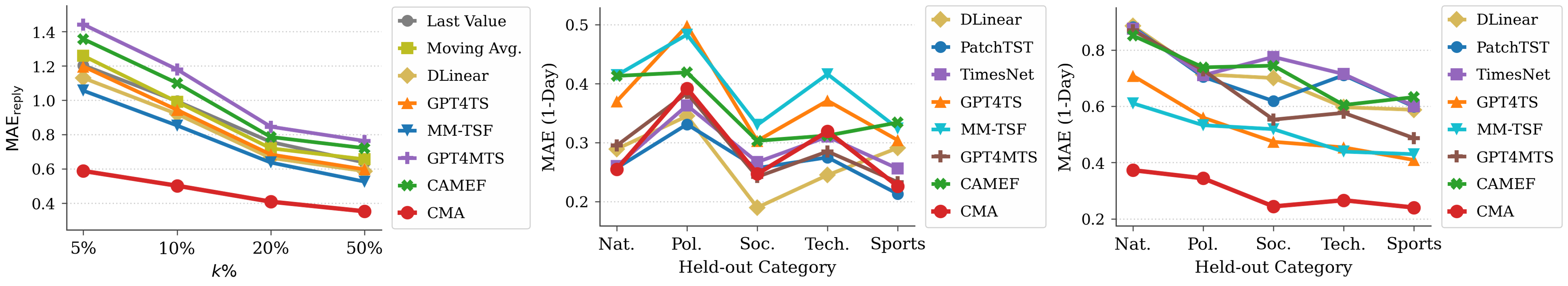}
\caption{Structure-aware and cross-category evaluation. \textbf{Left:} the variation of structure-aware $\mathrm{MAE}_{\mathrm{reply}}(k\%)$ across models on Sentiment Polarity at 1D as the interaction-density threshold $k\%$ changes. \textbf{Middle:} leave-one-category-out cross-event MAE on Discussion Intensity at 1D for each held-out category (Natural Disasters, Political Events, Social Movements, Technology Releases, and Sports and Entertainment Events). \textbf{Right:} leave-one-category-out cross-event MAE on Sentiment Polarity at 1D for each held-out category.}
\label{fig:mae_reply_cross_event}
\end{figure*}

\vspace{-0.2cm}
\section{Experiment Settings}
\label{sec:exp_setup}

Whereas Section~\ref{sec:benchmark_design} formalizes the forecasting task and the multimodal input space on SURGE, this section operationalizes the experimental setup that turns this formalization into concrete comparisons across baselines and configurations.

\vspace{-0.2cm}
\paragraph{Forecasting Setup.}
The experiments instantiate the forecasting task formalized in Section~\ref{sec:benchmark_design} across event-level numerical-only modeling, text-augmented modeling, structure-aware evaluation, and cross-category generalization.
The numerical setting uses only the historical time series $\mathbf{x}_{1:L}^{(e)}$ as input.
The text-augmented setting additionally conditions on the aligned textual observations $\mathbf{t}_{1:L}^{(e)}$ and is instantiated in three configurations: a no-text configuration that drops $\mathbf{t}$ entirely, a flat configuration that uses the linearized view, and a structured configuration that uses the reply-chain hierarchy. This three-way comparison exposes how each multimodal baseline responds to textual content versus reply-chain structure.
Each event is split chronologically into 70\% training, 10\% validation, and 20\% test segments.
For cross-category generalization, we adopt the leave-one-category-out protocol introduced in Section~\ref{sec:benchmark_design}, where all events from one category are held out for testing and the remaining categories are used for training.

\vspace{-0.2cm}
\paragraph{Metrics.}
We use mean absolute error (MAE) and mean squared error (MSE) as overall error indicators.
Both weight all time steps equally, whereas highly interactive periods on social media often coincide with sharper and more consequential event dynamics that holistic errors can obscure.
To evaluate forecasting performance on structurally dense periods, we therefore introduce a structure-aware metric, $\mathrm{MAE}_{\mathrm{reply}}(k\%)$, that computes MAE only on the most reply-dense bins.
For each temporal bin $t$, let $r_t$ denote the reply ratio, namely the fraction of posts in $\mathcal{P}_t$ that participate in a reply relationship:
\begin{equation}
r_t \;=\; \frac{\bigl|\{p \in \mathcal{P}_t : p \text{ is a reply or receives at least one reply}\}\bigr|}{|\mathcal{P}_t|}.
\label{eq:reply_ratio}
\end{equation}
Given a threshold $k\%$, the high-interaction subset $\mathcal{S}_{\mathrm{reply}}(k\%)$ collects the test bins whose reply ratios rank in the top $k\%$ when pooled globally across all test bins of all events at the given granularity:
\begin{equation}
\mathcal{S}_{\mathrm{reply}}(k\%) \;=\; \bigl\{t \in \mathcal{T}_{\mathrm{test}} \;:\; r_t \geq \mathrm{Percentile}_{100-k}\bigl(\{r_\tau\}_{\tau \in \mathcal{T}_{\mathrm{test}}}\bigr)\bigr\}.
\end{equation}
the structure-aware metric is then computed as the MAE over this subset:
\begin{equation}
\mathrm{MAE}_{\mathrm{reply}}(k\%) \;=\; \frac{1}{|\mathcal{S}_{\mathrm{reply}}(k\%)|}\sum_{t \in \mathcal{S}_{\mathrm{reply}}(k\%)} |x_t-\hat{x}_t|.
\end{equation}
We report this metric for $k \in \{5, 10, 20, 50\}$, which exposes behavioral differences on highly interactive and volatile time steps that are not visible from holistic MAE alone.

\vspace{-0.2cm}
\paragraph{Numerical baselines.}
For numerical-only forecasting we benchmark eight learned TSF models from the linear (DLinear \cite{zeng2023dlinear}), Transformer-based (PatchTST \cite{nie2023patchtst}, iTransformer \cite{liu2024itransformer}), convolutional (TimesNet \cite{wu2023timesnet}, ConvTimeNet \cite{cheng2024convtimenet}), LLM-based (GPT4TS \cite{zhou2024one}), and diffusion-based (TimeDiff \cite{shen2023timediff}, NsDiff \cite{fan2024nsdiff}) families \cite{lim2021tssurvey, qiu2024tfb,su2025diffusion}, alongside Last Value and Moving Average as naive reference floors.

\vspace{-0.2cm}
\paragraph{Text-augmented baselines.}
For text-augmented forecasting we adopt three multimodal TSF models originally designed for news- or report-paired numerical signals: MM-TSF \cite{anonymous2024mmtsflib}, GPT4MTS \cite{jia2024gpt4mts}, and CAMEF \cite{liu2024camef}. Each is evaluated under three input configurations (no-text, flat text, structured text), so that within-baseline deltas isolate the contribution of textual content (no-text to flat) and of reply-chain structure (flat to structured).

\vspace{-0.2cm}
\paragraph{Reference structure-aware probe.}
We additionally include cross-modal attention (CMA) as a lightweight reference probe that consumes per-post type and thread embeddings via intra-bin self-attention before fusing with the time-series encoder (Appendix~\ref{sec:appendix_cma}). We report it for diagnostic comparison rather than as a methodological contribution.

\vspace{-0.2cm}
\paragraph{Implementation details.}
Numerical-only experiments are conducted at three temporal granularities, namely 1-day, 12-hour, and 6-hour, with a consistent lookback-to-horizon ratio: $L{=}14, H{=}7$ for 1-day, $L{=}28, H{=}14$ for 12-hour, and $L{=}56, H{=}28$ for 6-hour. The text-augmented, structure-aware, and cross-category experiments are reported at the 1-day granularity with $L{=}14, H{=}7$ to keep the multimodal compute tractable. All TSF models are run with their official implementations, all reported MAE and MSE values are computed on the per-event z-score scale defined in Appendix~\ref{sec:appendix_ts_construction}, and all reported results are averaged over five random seeds. Per-cell across-seed standard deviations are reported in Appendix~\ref{sec:appendix_seed_std}.
See Appendix~\ref{sec:appendix_models} for per-model configurations and training setups.

\section{Results and Analyses}
\label{sec:results_and_analyses}
% ============================================================
\subsection{Numerical Forecasting Results}
As shown in the left panel of Figure~\ref{fig:num_text_results}, MAE and MSE favor different model families across both targets. Naive baselines, including Last Value and Moving Average, rank at the top under MAE on Discussion Intensity and Sentiment Polarity at nearly all granularities, whereas learned models more frequently rank at the top under MSE.
This separation indicates strong local persistence in event-driven series, where the best short-horizon forecast stays close to the most recent observation, while learned models trade larger average pointwise deviation for fewer extreme misses, producing smoother and more conservative predictions that are less prone to the outliers that inflate squared error.
Naive baselines remain competitive at 1-day granularity on both targets, while at 12-hour and 6-hour granularities GPT4TS additionally matches or surpasses them on Discussion Intensity. Finer granularities also lower absolute MAE on both targets. These trends indicate that the dominant challenge is the coexistence of strong local persistence and occasional sharp deviations rather than temporal resolution alone.

\vspace{-0.2cm}
\subsection{Text-Augmented Forecasting Results}
If reply-chain structure carries forecasting-relevant signal, multimodal inputs should rank above numerical-only models on both targets and the structured view should rank above the flat view within each baseline.
The right panel of Figure~\ref{fig:num_text_results} shows that among the text-only multimodal baselines only MM-TSF improves monotonically from no-text to flat to structured on both targets, while GPT4MTS is largely insensitive to either textual content or structural form and CAMEF responds only weakly and inconsistently across targets. Off-the-shelf multimodal TSF architectures designed for news-paired numerical signals therefore do not transfer directly to event-driven social-media data.
The CMA reference probe attains the lowest MAE on both targets under the structured configuration, with a markedly larger margin over text-blind baselines on Sentiment Polarity than on Discussion Intensity. Together, the within-baseline flat-to-structured deltas and the CMA contrast indicate that the preserved reply structure carries signal recoverable both through input formatting and through architecture-level consumption of typed post tokens.

\vspace{-0.2cm}
\subsection{Structure-Aware Evaluation on Sentiment Polarity}
The left panel of Figure~\ref{fig:mae_reply_cross_event} shows that every model's $\mathrm{MAE}_{\mathrm{reply}}(k\%)$ rises as $k$ decreases from 50\% to 5\%, indicating that reply-dense bins form an intrinsically harder slice of the benchmark independent of any one architecture.
Model rankings reshuffle as $k$ tightens: models that lead under holistic MAE need not lead in the densest interaction slices, while CMA-class models maintain a visible gap over text-blind baselines across the full range of $k$.
The three CMA configurations (no-text, flat, structured) remain close on this slice.
Because Sentiment Polarity is not monotonic in post volume, the high-interaction subset is not a volume-confounded easier-or-harder split.

\vspace{-0.2cm}
\subsection{Cross-Category Generalization}
On Sentiment Polarity (right panel of Figure~\ref{fig:mae_reply_cross_event}), the CMA reference probe ranks lowest on every one of the five held-out categories.
Cross-category difficulty varies clearly by held-out category, with natural disaster the hardest for most models and social movement the easiest.
On Discussion Intensity (middle panel) the ordering reverses: PatchTST attains the lowest average MAE across held-out categories, with DLinear close behind, while higher-capacity Transformer-based baselines lead on Sentiment Polarity.
These phenomena indicate that SURGE separates models suited to within-event forecasting from those that transfer more reliably across unseen categories, and that the two targets exercise distinct modeling capacities under the same protocol.
\section{Discussion}
\label{sec:discussion}
SURGE is a stress test for event-driven social-media forecasting. It exposes three benchmark properties, detailed below.

\vspace{-0.2cm}
\paragraph{Finding 1: persistence is strong but insufficient.}
Generic TSF~\cite{ansari2024chronos, das2024timesfm, woo2024moirai, gruver2024llmtime, zhou2021informer, wu2021autoformer, zhou2022fedformer, zhang2023crossformer, wang2024timemixer, liu2022nonstationary, lim2021tft} is typically designed around stable sources with physically regular dynamics such as sensor or meteorological data, whereas SURGE is driven by the interplay of event development, social organization, opinion steering, and information diffusion, which has no counterpart in standard benchmarks. This mismatch shows as an optimization-evaluation misalignment: learned time series models reduce MSE but fail to improve MAE over the Last Value baseline, indicating that squared-error objectives suppress large deviations without translating into central-tendency accuracy on event-driven series. Persistence is therefore a strong but insufficient baseline, and two distinct progress axes follow: improving central-tendency accuracy beyond persistence under MAE, and handling the rare extreme deviations that drive squared error.

\vspace{-0.2cm}
\paragraph{Finding 2: text is not automatically useful.}
Multimodal TSF models that jointly exploit historical numerical signal and aligned text only partially close the gap, and the benefit is uneven across baselines. Existing multimodal baselines are designed on textual side information whose internal structure does not itself encode interaction. Financial news paired with asset price series, for which methods such as GPT4MTS and CAMEF are originally designed, treats text as a unidirectional authoritative signal with no addressee structure, contrasting sharply with social media data that exhibit user-to-user reply chains \cite{tian2024learning,gan2026reinforced}. The benchmark therefore distinguishes whether a multimodal model handles social-media textual input at all from whether it handles it well.

\vspace{-0.2cm}
\paragraph{Finding 3: interaction density changes the evaluation regime.}
Aggregate MAE averages across sparse and dense interaction periods, obscuring where models actually fail. The structure-aware $\mathrm{MAE}_{\mathrm{reply}}(k\%)$ metric exposes that reply-dense bins form an intrinsically harder slice and that model rankings on this slice diverge from rankings under aggregate MAE. Cross-event heterogeneity adds a second axis along the same direction: leave-one-category-out transfer is structurally asymmetric~\cite{kamarthi2024crossdomainTS, liu2024tsood,liu2025balanced,zhang2026training,}, with exogenously driven categories such as natural disasters being the hardest to generalize to, so that improvements within a single event do not automatically carry across event types. SURGE therefore makes both interaction density and category shift available as benchmark axes along which existing methods fall short. Both axes carry application weight: in crisis response and opinion monitoring, the high-interaction periods and the categories that fall outside training distributions are precisely the cases where forecasting matters most, and where aggregate metrics tend to overestimate model readiness.

\vspace{-0.2cm}
\section{Limitations}
\label{sec:limitations}

SURGE has four limitations.
First, SURGE draws from Twitter, Reddit, and Threads and retains English text only, reflecting the dominant language of accessible content on these platforms and the validation overhead a multi-lingual annotation pipeline would entail. Communities centered on Chinese-language discussion (e.g., Weibo), short-video and image-first sharing (e.g., TikTok, Instagram), and semi-private group messaging (e.g., Telegram) fall outside SURGE's coverage, and events whose discussion concentrates on those venues may be under-represented or absent. Future work can extend SURGE along both the language and platform axes under comparable curation and structure-preservation procedures.
Second, the benchmark focuses on the forecasting task. Other tasks such as event classification, rumor detection, or stance shift analysis require additional annotations beyond what we provide and are left to future work.
Third, sentiment annotation is produced by a general-purpose LLM without domain-specific fine-tuning, and uses a three-class scheme rather than finer-grained five-level polarity or emotion categories because the latter carry higher inter-annotator disagreement and yield less stable automated labels at the event-aggregate level. Future work can build a domain-tuned sentiment model from SURGE's post-level text and extend to finer-grained public emotion forecasting by leveraging richer label schemes at lower aggregation levels.
Fourth, the released Sentiment Polarity captures expressed affect rather than stance, so conflict-related analyses should pair it with stance-specific or domain-tuned validators before any policy-relevant claim (Appendix~\ref{sec:appendix_ethics}).
% \vspace{-0.1cm}
\section{Conclusion}
\label{sec:conclusion}

In this paper, we present SURGE, an event-centric social media benchmark that organizes public discussion into multi-granularity time series paired with aligned text and reply/repost interaction structure.
Rather than proposing a new forecasting model as the main contribution, SURGE provides data, evaluation protocols, and diagnostic measurements for studying event-driven social-media forecasting.
Benchmark results indicate that the task is governed by strong local persistence, that existing text-augmented forecasters transfer unevenly to social media discussion, and that reply-dense periods constitute a harder evaluation slice than aggregate metrics reveal.
The included structure-aware probe serves as a reference implementation demonstrating how the released interaction fields can be consumed, while leaving the design of stronger interaction-aware forecasters to future work.

\bibliographystyle{unsrtnat}
\bibliography{main}

%%%%%%%%%%%%%%%%%%%%%%%%%%%%%%%%%%%%%%%%%%%%%%%%%%%%%%%%%%%%
\appendix

%%%%%%%%%%%%%%%%%%%%%%%%%%%%%%%%%%%%%%%%%%%%%%%%%%%%%%%%%%%%

% 附录 A：事件列表
\section{Event List}
\label{sec:appendix_events}

Table~\ref{tab:event_list} lists the full set of 67 events in SURGE, grouped by the five event categories discussed in Section~\ref{sec:data_source_processing}. The per-category counts are: Natural Disaster (12), Political (17), Social Movement (12), Technology (12), and Sports \& Entertainment (14).

\begin{table}[h]
\centering
\caption{Events in SURGE grouped by category. The 67 events span 2022--2026 and cover five distinct categories.}
\label{tab:event_list}
\begin{tabular}{p{0.22\textwidth} p{0.72\textwidth}}
\toprule
\textbf{Category (count)} & \textbf{Events} \\
\midrule

Natural Disaster (12) &
Baltimore Bridge Collapse;
California Wildfires 2025;
Europe Heatwave \& Wildfire;
Hurricane Fiona;
Hurricane Ian;
Indian Plane Crash;
Noto Earthquake (Japan);
Pakistan Floods 2022;
Table Rock (SC) Fire;
Texas Flood;
Turkey--Syria Earthquake;
US Wildfire \\

\addlinespace
Political (17) &
20th CPC National Congress;
Battle of Bakhmut 2023;
Damascus--SDF Conflict;
Gaza Ceasefire;
India General Election 2024;
India--Pakistan Conflict;
Iran--Israel Conflict;
Israel--Hamas Conflict 2025;
Israeli--Palestinian Conflict 2023;
Kazakhstan Protests 2022;
Rafah Military Operation;
TikTok Ban;
Trump Assassination Attempt;
Trump Global Tariff;
Trump Inauguration;
Russia--Ukraine Conflict 2022;
US Government Shutdown 2025 \\

\addlinespace
Social Movement (12) &
Anne Burrell Event;
Brigitte Bardot Death 2025;
Charlie Kirk Event;
Israel Judicial Reform Protests;
Kanye West Controversy;
Kate Middleton Controversy;
Moscow Concert Hall Incident;
Queen Elizabeth II Death;
Sean `Diddy' Combs Investigation;
Sri Lanka Protests;
Third September Military Parade;
Will Smith Slap 2022 \\

\addlinespace
Technology (12) &
AI Multimodal Launch;
Artemis Moon Mission;
Axiom Mission 2022;
Axiom Mission 2023;
Chang'e 6 Moon Mission;
DeepSeek V3.x Releases;
Google Gemini Release;
GPT Release;
GPT-5.3 \& Codex Release;
James Webb Space Telescope;
New Glenn Milestone;
Robotaxi Launch \\

\addlinespace
Sports \& Entertainment (14) &
American Baseball League;
Australian Open 2026;
Beijing Winter Olympics 2022;
Cannes Film Festival 2024;
FIFA Club World Cup 2025;
FIFA Women's World Cup;
FIFA World Cup 2022;
Golden Globes 2026;
Grammys 2026;
Gwyneth Paltrow Trial;
NBA Finals;
Paris Olympics 2024;
Super Bowl LX 2026;
T20 World Cup 2026 Final \\

\bottomrule
\end{tabular}
\end{table}

\subsection{Per-Event Metadata}
\label{sec:appendix_event_metadata}

Tables~\ref{tab:event_metadata_nd}--\ref{tab:event_metadata_se} provide per-event collection and processing statistics for the 67 events listed above, grouped into one table per event category. The Window column reports the calendar span between the earliest and latest retained post timestamps. Platform shares (T = Twitter, R = Reddit, Th = Threads) are reported as percentages of retained posts and rounded to integers. The Raw column gives the post count returned by the acquisition queries. Filt.\ gives the count after post-level filtering (Appendix~\ref{sec:appendix_schema_filtering}). Active gives the count inside the detected active period (Appendix~\ref{sec:appendix_ts_construction}). The three Bins columns report the active-period bin count at each granularity. Events whose active-period bin count at a given granularity falls below the minimum-bin requirement of 21 are excluded from that granularity's benchmark splits while still being released as time series. The per-granularity coverage of $55$, $64$, and $67$ events at 1-day, 12-hour, and 6-hour reflects this exclusion.

\begin{table}[h]
\centering
\caption{Per-event metadata: Natural Disaster events.}
\label{tab:event_metadata_nd}
\scriptsize
\setlength{\tabcolsep}{3.5pt}
\begin{tabular}{l c c r r r r r r}
\toprule
\textbf{Event} & \textbf{Window} & \textbf{T:R:Th} & \textbf{Raw} & \textbf{Filt.} & \textbf{Active} & \textbf{1D} & \textbf{12H} & \textbf{6H} \\
\midrule
Baltimore Bridge Collapse  & 2024/03--2024/04 & 100:0:0 & 2{,}983 & 2{,}510 & 2{,}506 & 8 & 15 & 31 \\
California Wildfires 2025  & 2024/12--2026/01 & 40:60:0 & 29{,}640 & 26{,}288 & 26{,}288 & 110 & 448 & 1{,}650 \\
Europe Heatwave \& Wildfire & 2022/06--2025/08 & 97:3:0 & 45{,}982 & 41{,}332 & 41{,}332 & 420 & 2{,}332 & 4{,}663 \\
Hurricane Fiona            & 2022/09--2023/06 & 100:0:0 & 4{,}835 & 4{,}284 & 3{,}605 & 12 & 24 & 48 \\
Hurricane Ian              & 2022/09--2024/09 & 100:0:0 & 8{,}174 & 7{,}106 & 3{,}302 & 7 & 13 & 26 \\
Indian Plane Crash         & 2025/06--2025/07 & 100:0:0 & 2{,}546 & 1{,}067 & 1{,}067 & 35 & 70 & 138 \\
Noto Earthquake (Japan)    & 2023/12--2024/06 & 53:47:0 & 4{,}147 & 2{,}731 & 2{,}713 & 24 & 62 & 123 \\
Pakistan Floods 2022       & 2022/08--2024/04 & 100:0:0 & 5{,}765 & 4{,}394 & 4{,}363 & 42 & 83 & 165 \\
Table Rock (SC) Fire       & 2025/03--2025/04 & 100:0:0 & 474 & 403 & 403 & 21 & 41 & 81 \\
Texas Flood                & 2025/07--2025/11 & 81:0:19 & 6{,}280 & 3{,}290 & 3{,}078 & 12 & 24 & 48 \\
Turkey--Syria Earthquake   & 2023/02--2025/01 & 28:72:0 & 6{,}409 & 4{,}607 & 4{,}591 & 49 & 97 & 223 \\
US Wildfire                & 2025/01--2025/05 & 100:0:0 & 5{,}614 & 2{,}561 & 2{,}559 & 19 & 36 & 219 \\
\bottomrule
\end{tabular}
\end{table}

\begin{table}[h]
\centering
\caption{Per-event metadata: Political events.}
\label{tab:event_metadata_pol}
\scriptsize
\setlength{\tabcolsep}{3.5pt}
\begin{tabular}{l c c r r r r r r}
\toprule
\textbf{Event} & \textbf{Window} & \textbf{T:R:Th} & \textbf{Raw} & \textbf{Filt.} & \textbf{Active} & \textbf{1D} & \textbf{12H} & \textbf{6H} \\
\midrule
20th CPC National Congress    & 2022/10--2022/11 & 100:0:0 & 232 & 133 & 133 & 24 & 46 & 92 \\
Battle of Bakhmut 2023        & 2023/01--2023/03 & 0:100:0 & 1{,}030 & 910 & 910 & 88 & 175 & 349 \\
Damascus--SDF Conflict        & 2025/12--2026/03 & 100:0:0 & 1{,}511 & 1{,}212 & 1{,}212 & 90 & 179 & 356 \\
Gaza Ceasefire                & 2024/12--2025/12 & 45:55:0 & 8{,}656 & 7{,}503 & 7{,}503 & 238 & 746 & 1{,}491 \\
India General Election 2024   & 2024/03--2024/06 & 100:0:0 & 2{,}332 & 1{,}487 & 1{,}487 & 107 & 213 & 424 \\
India--Pakistan Conflict      & 2025/04--2025/08 & 77:23:0 & 25{,}325 & 18{,}065 & 17{,}948 & 23 & 45 & 90 \\
Iran--Israel Conflict         & 2025/06--2025/12 & 29:71:0 & 8{,}212 & 5{,}110 & 4{,}891 & 18 & 36 & 72 \\
Israel--Hamas Conflict 2025   & 2025/09--2026/03 & 100:0:0 & 81{,}720 & 66{,}649 & 66{,}628 & 57 & 114 & 228 \\
Israeli--Palestinian Conflict 2023 & 2023/10--2025/12 & 35:65:0 & 14{,}892 & 12{,}238 & 10{,}978 & 82 & 163 & 331 \\
Kazakhstan Protests 2022      & 2021/07--2022/03 & 100:0:0 & 2{,}026 & 1{,}264 & 1{,}214 & 30 & 62 & 123 \\
Rafah Military Operation      & 2024/02--2025/12 & 100:0:0 & 24{,}837 & 14{,}478 & 14{,}478 & 421 & 1{,}348 & 2{,}694 \\
TikTok Ban                    & 2024/05--2025/11 & 83:0:17 & 1{,}372 & 573 & 467 & 3 & 5 & 31 \\
Trump Assassination Attempt   & 2024/06--2025/12 & 79:21:0 & 22{,}809 & 19{,}874 & 19{,}716 & 50 & 99 & 201 \\
Trump Global Tariff           & 2024/11--2025/07 & 84:16:0 & 23{,}958 & 19{,}816 & 19{,}371 & 45 & 92 & 183 \\
Trump Inauguration            & 2024/12--2026/01 & 29:71:0 & 44{,}639 & 38{,}908 & 38{,}761 & 85 & 177 & 356 \\
Russia--Ukraine Conflict 2022 & 2018/02--2025/06 & 100:0:0 & 8{,}217 & 5{,}436 & 5{,}215 & 55 & 109 & 221 \\
US Government Shutdown 2025   & 2025/09--2026/03 & 100:0:0 & 106{,}709 & 89{,}647 & 89{,}561 & 72 & 143 & 285 \\
\bottomrule
\end{tabular}
\end{table}

\begin{table}[h]
\centering
\caption{Per-event metadata: Social Movement events.}
\label{tab:event_metadata_sm}
\scriptsize
\setlength{\tabcolsep}{3.5pt}
\begin{tabular}{l c c r r r r r r}
\toprule
\textbf{Event} & \textbf{Window} & \textbf{T:R:Th} & \textbf{Raw} & \textbf{Filt.} & \textbf{Active} & \textbf{1D} & \textbf{12H} & \textbf{6H} \\
\midrule
Anne Burrell Event             & 2025/01--2025/11 & 77:0:23 & 5{,}213 & 1{,}770 & 1{,}342 & 13 & 26 & 55 \\
Brigitte Bardot Death 2025     & 2025/12--2026/01 & 100:0:0 & 1{,}633 & 1{,}456 & 1{,}456 & 43 & 86 & 172 \\
Charlie Kirk Event             & 2025/09--2025/10 & 54:46:0 & 10{,}342 & 8{,}109 & 7{,}994 & 17 & 34 & 67 \\
Israel Judicial Reform Protests & 2023/03--2023/05 & 0:100:0 & 698 & 643 & 643 & 64 & 126 & 252 \\
Kanye West Controversy         & 2022/10--2025/09 & 100:0:0 & 8{,}811 & 7{,}836 & 7{,}780 & 33 & 80 & 200 \\
Kate Middleton Controversy     & 2024/02--2024/06 & 100:0:0 & 1{,}855 & 1{,}419 & 1{,}419 & 100 & 198 & 395 \\
Moscow Concert Hall Incident   & 2024/03--2024/04 & 100:0:0 & 2{,}134 & 1{,}064 & 1{,}064 & 11 & 21 & 44 \\
Queen Elizabeth II Death       & 2022/09--2025/03 & 100:0:0 & 9{,}312 & 8{,}663 & 8{,}605 & 62 & 123 & 246 \\
Sean `Diddy' Combs Investigation & 2024/03--2025/12 & 74:26:0 & 16{,}545 & 13{,}421 & 13{,}421 & 660 & 1{,}320 & 2{,}640 \\
Sri Lanka Protests             & 2022/06--2025/11 & 100:0:0 & 4{,}747 & 3{,}968 & 3{,}962 & 31 & 75 & 221 \\
Third September Military Parade & 2025/06--2025/11 & 90:5:5 & 23{,}217 & 10{,}464 & 9{,}877 & 29 & 57 & 112 \\
Will Smith Slap 2022           & 2022/03--2023/01 & 100:0:0 & 2{,}449 & 2{,}132 & 2{,}132 & 303 & 605 & 1{,}209 \\
\bottomrule
\end{tabular}
\end{table}

\begin{table}[h]
\centering
\caption{Per-event metadata: Technology events.}
\label{tab:event_metadata_tech}
\scriptsize
\setlength{\tabcolsep}{3.5pt}
\begin{tabular}{l c c r r r r r r}
\toprule
\textbf{Event} & \textbf{Window} & \textbf{T:R:Th} & \textbf{Raw} & \textbf{Filt.} & \textbf{Active} & \textbf{1D} & \textbf{12H} & \textbf{6H} \\
\midrule
AI Multimodal Launch       & 2025/04--2026/02 & 100:0:0 & 2{,}026 & 1{,}384 & 1{,}384 & 314 & 627 & 1{,}254 \\
Artemis Moon Mission       & 2022/11--2024/12 & 100:0:0 & 2{,}035 & 1{,}850 & 1{,}843 & 42 & 98 & 196 \\
Axiom Mission 2022         & 2022/04--2025/03 & 100:0:0 & 2{,}442 & 1{,}776 & 1{,}050 & 61 & 120 & 239 \\
Axiom Mission 2023         & 2023/05--2023/06 & 100:0:0 & 777 & 684 & 684 & 54 & 107 & 212 \\
Chang'e 6 Moon Mission     & 2024/05 & 100:0:0 & 247 & 75 & 75 & 24 & 47 & 93 \\
DeepSeek V3.x Releases     & 2025/01--2026/01 & 13:87:0 & 31{,}588 & 25{,}724 & 10{,}545 & 89 & 184 & 368 \\
Google Gemini Release      & 2024/01--2026/01 & 100:0:0 & 9{,}740 & 7{,}351 & 7{,}351 & 725 & 1{,}450 & 2{,}899 \\
GPT Release                & 2023/03--2026/01 & 17:82:1 & 49{,}397 & 28{,}977 & 28{,}421 & 365 & 730 & 1{,}460 \\
GPT-5.3 \& Codex Release   & 2026/02--2026/03 & 100:0:0 & 912 & 869 & 869 & 42 & 83 & 165 \\
James Webb Space Telescope & 2022/07--2022/09 & 100:0:0 & 1{,}806 & 1{,}615 & 1{,}614 & 33 & 66 & 133 \\
New Glenn Milestone        & 2025/11--2026/01 & 100:0:0 & 1{,}152 & 1{,}013 & 1{,}013 & 75 & 149 & 297 \\
Robotaxi Launch            & 2024/09--2026/01 & 13:86:1 & 56{,}007 & 45{,}077 & 30{,}578 & 153 & 306 & 612 \\
\bottomrule
\end{tabular}
\end{table}

\begin{table}[h]
\centering
\caption{Per-event metadata: Sports \& Entertainment events.}
\label{tab:event_metadata_se}
\scriptsize
\setlength{\tabcolsep}{3.5pt}
\begin{tabular}{l c c r r r r r r}
\toprule
\textbf{Event} & \textbf{Window} & \textbf{T:R:Th} & \textbf{Raw} & \textbf{Filt.} & \textbf{Active} & \textbf{1D} & \textbf{12H} & \textbf{6H} \\
\midrule
American Baseball League   & 2024/10--2024/11 & 100:0:0 & 4{,}303 & 1{,}540 & 1{,}540 & 16 & 32 & 87 \\
Australian Open 2026       & 2026/01--2026/02 & 100:0:0 & 10{,}642 & 9{,}099 & 9{,}096 & 31 & 62 & 125 \\
Beijing Winter Olympics 2022 & 2022/01--2025/02 & 80:20:0 & 19{,}952 & 16{,}054 & 15{,}849 & 53 & 105 & 209 \\
Cannes Film Festival 2024  & 2024/04--2024/09 & 100:0:0 & 1{,}783 & 901 & 901 & 144 & 288 & 575 \\
FIFA Club World Cup 2025   & 2025/06--2025/11 & 96:0:4 & 42{,}971 & 10{,}324 & 10{,}323 & 146 & 308 & 615 \\
FIFA Women's World Cup     & 2023/07--2023/12 & 100:0:0 & 3{,}371 & 2{,}580 & 2{,}580 & 161 & 322 & 644 \\
FIFA World Cup 2022        & 2022/11--2026/01 & 100:0:0 & 30{,}443 & 22{,}202 & 22{,}202 & 1{,}174 & 2{,}348 & 4{,}696 \\
Golden Globes 2026         & 2025/12--2026/01 & 100:0:0 & 1{,}209 & 1{,}108 & 1{,}108 & 54 & 108 & 216 \\
Grammys 2026               & 2025/11--2026/03 & 100:0:0 & 11{,}042 & 9{,}125 & 9{,}125 & 115 & 230 & 519 \\
Gwyneth Paltrow Trial      & 2023/03--2023/04 & 0:100:0 & 321 & 290 & 290 & 23 & 45 & 89 \\
NBA Finals                 & 2024/10--2025/11 & 98:1:1 & 259{,}982 & 86{,}462 & 85{,}959 & 254 & 508 & 1{,}017 \\
Paris Olympics 2024        & 2024/06--2025/12 & 23:77:0 & 91{,}279 & 54{,}652 & 54{,}647 & 122 & 918 & 2{,}059 \\
Super Bowl LX 2026         & 2026/01--2026/03 & 100:0:0 & 23{,}542 & 19{,}988 & 19{,}966 & 26 & 51 & 102 \\
T20 World Cup 2026 Final   & 2026/02--2026/03 & 100:0:0 & 2{,}941 & 2{,}585 & 2{,}584 & 18 & 36 & 71 \\
\bottomrule
\end{tabular}
\end{table}

% 附录 B：数据收集细节
%
\section{Data Collection Details}
\label{sec:appendix_data_collection}

This appendix supplements Section~\ref{sec:data_collection} with procedural details of the raw collection from which SURGE is constructed.

\paragraph{Acquisition Campaigns.}
The 1{,}256{,}816 raw posts cover events that occurred between 2022 and 2026 and were gathered through multiple independent acquisition campaigns, each targeting a specific event or topical cluster within one of the five event categories.
For each campaign, candidate query terms combine the canonical event name with co-occurring entities, hashtags, and key phrases identified from initial probe queries to broaden coverage of the event's discourse.

\paragraph{Schema Heterogeneity.}
Because the campaigns were organized independently and across different time periods, the raw records arrive in heterogeneous formats with inconsistent field naming, varying metadata completeness, and platform-specific encoding conventions.
For example, Twitter records distinguish quote posts from replies through dedicated metadata fields, whereas Reddit records use a parent-link convention that requires additional traversal to recover the same relation.
A unified post schema with standardized timestamps, textual content, engagement statistics, platform identity, and provenance information therefore precedes all subsequent processing.

\paragraph{Candidate Event Pool.}
Raw collection produced 93 candidate events, with each category over-sampled relative to its final target count to absorb downstream filtering loss.
Events removed during processing failed one or more criteria specified in Appendix~\ref{sec:appendix_data_preprocessing} and Appendix~\ref{sec:appendix_ts_construction}, in particular insufficient active-period length and post volume below the threshold required to instantiate a valid forecasting window.

% 附录 C：数据预处理细节（含情感标注）
%
\section{Data Preprocessing Details}
\label{sec:appendix_data_preprocessing}

This appendix provides procedural details of the preprocessing pipeline summarized in Section~\ref{sec:collection_filtering}.

\subsection{Schema Unification and Quality Filtering}
\label{sec:appendix_schema_filtering}

All raw records are first mapped to the unified post schema introduced in Appendix~\ref{sec:appendix_data_collection}, and posts collected across overlapping temporal windows or through redundant query expansions are deduplicated by matching post identifiers and content hashes.
The deduplicated pool then passes through two sequential filtering stages: post-level rules that remove individual records lacking sufficient text signal, and event-level thresholds that exclude events whose retained corpus cannot support reliable bin-level aggregation.

Post-level filtering applies five rules in a fixed priority order, with each post evaluated against the rules in turn and removed by the first matching rule. (i) Short-text removal discards records whose stripped text length is below 5 characters. (ii) Emoji-or-symbol-only removal discards records whose count of alphabetic characters after stripping Unicode emoji and punctuation is below 5, even when the raw character count is large. (iii) URL-spam removal discards records in which URL characters constitute more than 50\% of the total text length. (iv) Non-English-language removal discards records that the \texttt{langdetect} library\footnote{\url{https://github.com/Mimino666/langdetect}} identifies as non-English. This rule is applied only to texts of at least 20 characters after URL stripping so that short ambiguous texts are not removed on unreliable detector output. (v) Within-event textual deduplication discards records whose normalized text matches that of an earlier post in the same event, where normalization lowercases the text and strips URLs, \texttt{@mentions}, and redundant whitespace, and the first occurrence is kept.

Event-level filtering then removes events whose retained corpus is insufficient to instantiate a meaningful time series, applying three thresholds jointly: a minimum of 50 retained posts, a minimum time span of 3 days between the earliest and latest post timestamps, and a minimum post density of 3 posts per day on average. These thresholds are deliberately permissive, and most events that survive post-level filtering pass them with substantial margin. The tighter, event-relative active-period criterion applied within the surviving events is documented in Appendix~\ref{sec:appendix_ts_construction}.

\subsection{Sentiment Annotation Details}
\label{sec:appendix_sentiment}
Post-level sentiment labels are produced by Qwen3-32B.
The following zero-shot prompt is applied to every post without any domain-specific fine-tuning:
\begin{quote}
\small\ttfamily
Analyze the sentiment of the following social media comment.\\
Classify it as exactly one of: positive, neutral, negative.\\
Only output the single word classification, nothing else.\\[4pt]
Comment: \{text\}\\[4pt]
Sentiment:
\end{quote}
Resulting labels are mapped to numerical scores for time series construction, with positive corresponding to $+1$, neutral to $0$, and negative to $-1$.
Two considerations motivate the choice of a general-purpose LLM as the post-level annotator.
First, a domain-tuned classifier would itself constitute a separate research contribution and would introduce additional hyperparameters and design choices.
Second, the time series construction in SURGE aggregates tens to hundreds of posts per temporal bin, so individual labeling errors are substantially diluted at the event-aggregate level where all downstream analysis operates.
The choice of a three-class scheme rather than finer-grained five-level polarity or emotion-category taxonomies reflects the observation that finer schemes carry higher inter-annotator disagreement in human annotation studies and produce less stable automated labels, especially when aggregated at the event level.

\paragraph{Stratified Human Verification of LLM Labels.}
To assess the reliability of LLM-assigned sentiment labels and to detect category- or class-specific drift that an aggregate accuracy figure can hide, we conduct a human verification study on $3{,}000$ posts stratified jointly across the five event categories and the three sentiment classes, with $200$ posts per category-class cell drawn uniformly at random across events and platforms.
% 标注者与 codebook
Two graduate students with proficient English serve as annotators. They label the sample independently under a shared codebook that defines the three sentiment classes and resolves edge cases as follows: sarcasm and irony are labeled by intended rather than surface meaning, mixed-sentiment posts are labeled by the dominant attitude, and quoted content is labeled by the quoting author's stance toward the quoted material. Posts whose content lies outside the scope handled by upstream filtering are reported back to the authors and replaced from the sample pool.
The annotation interface displays the post text, the associated event name, and the timestamp, while concealing the LLM-assigned label and the user handle to prevent anchoring. A pilot round on $50$ posts disjoint from the verification set is used solely to refine the codebook, and pilot data are not included in the reported statistics.
% 度量与整体结果
After independent labeling, we compute Cohen's $\kappa$ between the two annotators on the raw labels, observing $\kappa = 0.74$ with an initial disagreement rate of $18.3\%$. Disagreements are resolved through joint discussion to produce a single consensus label per post. The LLM labels match the human consensus on $86.5\%$ of posts overall, with per-class F1 scores of $0.86$, $0.89$, and $0.82$ for positive, neutral, and negative sentiment respectively.
% 分层结果与 corpus-level bias 估计
Table~\ref{tab:sentiment_stratified} reports per-cell agreement within each category-class stratum, where the stratum is defined by the LLM-assigned label. Per-cell agreement remains within a narrow band, with the lowest agreement on negative posts in political events where sarcasm and stance ambiguity are most prevalent, and the highest on neutral posts in technology releases. Aggregating the signed misclassifications across the verification set into the corpus-level bias $\mu = \mathbb{E}[s_p^{\mathrm{LLM}} - s_p^{\star}]$, we obtain $|\mu| \leq 0.05$, which lies well below the residual MAE of the strongest baseline on Sentiment Polarity reported in Section~\ref{sec:exp_setup} and confirms that the bin-aggregate noise floor analysis below remains valid.

\begin{table}[h]
\centering
\caption{Stratified LLM-human agreement on the 3{,}000-post verification set ($200$ posts per category-class cell, with strata defined by the LLM-assigned label).}
\label{tab:sentiment_stratified}
\small
\begin{tabular}{l ccc}
\toprule
\multirow{2}{*}{\textbf{Category}}
& \multicolumn{3}{c}{\textbf{Per-class accuracy}} \\
\cmidrule(lr){2-4}
& Pos & Neu & Neg \\
\midrule
Natural Disaster        & 0.86 & 0.89 & 0.84 \\
Political               & 0.83 & 0.87 & 0.80 \\
Social Movement         & 0.85 & 0.88 & 0.82 \\
Technology              & 0.88 & 0.91 & 0.86 \\
Sports \& Entertainment & 0.87 & 0.90 & 0.85 \\
\bottomrule
\end{tabular}
\end{table}

\paragraph{Bin-Level Aggregation Bounds Per-Post Label Noise.}
The benchmark targets are bin-level aggregates rather than per-post labels, so the relevant validation question for the SURGE benchmark is not whether the LLM achieves high per-post accuracy but how much per-post labeling error survives bin-level aggregation. We give a precise statistical bound to formalize this relationship.
Let $s_p \in \{-1, 0, +1\}$ denote the LLM-assigned label for post $p$ and $s_p^\star$ the unobserved gold human label. Define the per-post labeling error as $\epsilon_p = s_p - s_p^\star$, and assume $|\epsilon_p| \leq 2$ since labels lie in $\{-1, 0, +1\}$. Suppose the LLM disagrees with the human label with probability at most $\alpha$ on average over the underlying post distribution, and write $\mu = \mathbb{E}[\epsilon_p]$ and $\sigma^2 = \mathrm{Var}[\epsilon_p]$ for the resulting bias and variance. Because $|\epsilon_p| \leq 2$ and $\epsilon_p = 0$ whenever the LLM and human labels agree, $\mathbb{E}[\epsilon_p^2] \leq 4\alpha$ and therefore $\sigma^2 \leq 4\alpha$.
The bin-level Sentiment Polarity error is
\begin{equation}
\bar{s}_t - \bar{s}_t^\star \;=\; \frac{1}{c_t}\sum_{p \in \mathcal{P}_t} \epsilon_p,
\end{equation}
which under the standard mild assumption that per-post errors within a bin are weakly dependent has expectation $\mu$ and variance bounded by $\kappa_t \sigma^2 / c_t$, where $\kappa_t \geq 1$ is a dependence inflation factor that equals one when errors are independent and remains a small constant under weak dependence. The random fluctuation around $\mu$ therefore has standard deviation $O(\sqrt{\kappa_t \alpha / c_t})$, and aggregation only attenuates this stochastic component while the systematic bias $\mu$ persists in the bin-level aggregate regardless of $c_t$. Across the released SURGE bins, active-period bins span roughly tens to hundreds of posts per bin in the typical regime, with event-level peak bins reaching $10^3$--$10^4$ posts. Plugging the bound $\sigma^2 \leq 4\alpha$ in under independent errors ($\kappa_t = 1$) and a per-post disagreement rate of $\alpha = 0.20$ gives a fluctuation bound of $\sqrt{0.8 / c_t}$ on the $[-1, +1]$ Sentiment Polarity scale, which falls below $0.09$ once $c_t \geq 100$ and below $0.03$ at peak bins with $c_t \geq 10^3$ posts. Reported MAE values in Section~\ref{sec:exp_setup} are computed on the per-event z-score scale, on which this raw-scale fluctuation is divided by the per-event standard deviation $\sigma_e$ of Sentiment Polarity. Even after this rescaling, the per-bin stochastic fluctuation remains substantially smaller than the residual MAE of strong baselines on the same z-score scale (e.g., $0.465$ for Last Value at 1-day granularity), so the stochastic component of bin-level annotation noise predicted by the bound at $\alpha=0.20$ is small relative to the residual error that benchmark models are trained to predict.

% 附录 D：时间序列构建细节
% 
\section{Time Series Construction Details}
\label{sec:appendix_ts_construction}

This appendix documents the procedural details of converting per-post sentiment labels into the released time series, the anonymized per-bin post-ID selections used to reconstruct text views, and the released anonymized interaction edges. It picks up from the post-level labels described in Appendix~\ref{sec:appendix_data_preprocessing} and covers binning, target derivation, active-period detection, missing-value handling, normalization, and text-view construction.

\paragraph{Calendar-Aligned Temporal Binning.}
For each event, we discretize its active period into fixed-width temporal bins at three granularities. Bin boundaries are aligned to absolute calendar time rather than to the event onset, ensuring that all events share a common temporal reference. Specifically, daily bins start at 00:00 Coordinated Universal Time (UTC), 12-hour bins start at either 00:00 or 12:00 UTC, and 6-hour bins start at one of 00:00, 06:00, 12:00, or 18:00 UTC. For a granularity $\Delta \in \{1\mathrm{d}, 12\mathrm{h}, 6\mathrm{h}\}$, each post with timestamp $t$ is assigned to the unique half-open interval $[t_{\mathrm{start}}, t_{\mathrm{start}}+\Delta)$ that contains $t$.

\paragraph{Active-Period Detection.}
Public events typically begin and end with long low-activity tails whose bin-level aggregates reflect incidental noise rather than genuine event dynamics, so a uniform absolute volume threshold would either trim sharp-burst events too aggressively or fail to trim slow-burn events at all. Each event is therefore trimmed using an event-relative criterion. Let $\bar{c}$ be the mean per-bin post count over the candidate timeline and let $\tau = 0.05 \cdot \bar{c}$. The active period is defined as the smallest contiguous interval $[t_L, t_R]$ such that $c_{t_L} \geq \tau$ and $c_{t_R} \geq \tau$, identified by scanning inward from each end until the first bin that crosses the threshold. Events whose active period contains fewer than 21 bins at a given granularity are excluded at that granularity. The threshold of 21 corresponds to the lookback-plus-horizon length $L+H$ at 1-day granularity (Section~\ref{sec:exp_setup}), so that the active period can in principle yield at least one forecasting window assigned to the test split under the chronological train/validation/test partitioning. This rule yields the asymmetric per-granularity coverage of 55 events at 1-day, 64 events at 12-hour, and 67 events at 6-hour.

\paragraph{Empty Bins and Missing-Value Imputation.}
Within an active period, bins may still contain zero posts and the released time series preserve this distinction by encoding $|\mathcal{P}_t| = 0$ as \texttt{NaN} in both target variables, so that downstream users retain the choice of imputation strategy.
The benchmark pipeline used in this paper imputes missing bins within each split independently. Each event is first divided chronologically into training, validation, and test segments, and forward fill is then applied within each segment to propagate the most recent in-segment observation across the gap. Any leading gap that remains in a segment because no earlier in-segment observation exists is filled with the first observation that appears later within the same segment, so no information ever crosses a split boundary. Within each segment, the forward fill step is strictly causal in the time-series sense, while the leading-gap backfill is not strictly causal but remains confined within its own segment and therefore introduces no cross-split leakage. This split-local imputation is selected over linear interpolation because interpolation would introduce a synthetic trend across silent intervals and inject information that the corresponding bins did not actually carry.

\paragraph{Per-Event Z-Score Normalization.}
For each event $e$ and each target variable, the z-score normalization
\begin{equation}
\tilde{x}_t \;=\; \frac{x_t - \mu_e}{\sigma_e},
\end{equation}
is applied, where $\mu_e$ and $\sigma_e$ are the mean and standard deviation computed on the training split of event $e$ only, and the same statistics are used to normalize the validation and test segments (variables with $\sigma_e = 0$ on the training split are mapped to the constant zero series). Per-event normalization is necessary because raw scales differ by orders of magnitude across events and across target variables (e.g., Discussion Intensity ranges from a handful of posts per bin on quieter events to $10^{3}$--$10^{4}$ posts per bin at the peak bins of high-volume events), and a global normalization would be dominated by the largest events and render MAE and MSE incomparable across the dataset. All MAE and MSE values reported in Section~\ref{sec:exp_setup} are computed on the resulting per-event z-score scale, so that errors aggregate comparably across events whose raw scales span several orders of magnitude.

\paragraph{Text View Construction.}
For each bin we define a fixed selection of posts that supports two textual views (flat and structured) sharing the same underlying content, so that comparisons across views isolate the effect of structure rather than of content. The selection is released as anonymized post IDs; users with platform access reconstruct the corresponding text views locally via the released hydration script that fetches post content from the originating platforms under their respective terms. SURGE does not redistribute raw or sampled post text. Within each bin, main posts are first ranked by the number of replies they receive within the same bin (used as a proxy for in-bin conversational engagement), and the top three main posts are retained. For each retained main post, up to two replies are then included in time order. If a bin contains no main post, the fallback is the three earliest posts in the bin in time order, each treated as a singleton thread. The selection set is identical for the structured view and the flat view, with the two views differing only in formatting.
The structured view organizes the selected posts into reply chains with explicit role markers, indenting each reply under its parent main post:
\begin{quote}
\small\ttfamily
UserA said: \{main-post text\}\\
\hphantom{Use}>> UserB replied: \{reply text\}\\
\hphantom{Use}>> UserC replied: \{reply text\}\\
UserD said: \{main-post text\}\\
\hphantom{Use}>> UserE replied: \{reply text\}
\end{quote}
The flat view linearizes the same posts in chronological order without role markers and without indentation:
\begin{quote}
\small\ttfamily
UserA: \{post text\}\\
UserB: \{post text\}\\
UserC: \{post text\}\\
UserD: \{post text\}\\
UserE: \{post text\}
\end{quote}

Both views are truncated to a maximum of 1{,}500 characters per bin, with the trailing ellipsis ``\texttt{...}'' appended whenever truncation occurs. This limit corresponds to the 512-token input budget of BERT \cite{devlin2019bert} and RoBERTa \cite{liu2019roberta}, which are the text encoders used by the multimodal baselines and by the CMA reference baseline.

% 附录 E：基线模型与实验设置
%
\section{Model Configurations and Experiment Setup}
\label{sec:appendix_models}

\subsection{Numerical Forecasting Models}

Table~\ref{tab:numerical_models} summarizes the ten numerical forecasting models evaluated in the numerical forecasting setting.
Two naive baselines establish lower bounds: Last Value repeats the final observed value across the entire prediction horizon, and Moving Average uses a trailing 7-step window.
DLinear decomposes the input into trend and residual components via a moving average kernel and applies separate linear projections to each.
PatchTST segments the input into non-overlapping patches and processes them with a Transformer encoder.
iTransformer inverts the standard Transformer by treating each time step as a token and applying attention across the temporal dimension.
TimesNet reshapes 1D time series into 2D tensors via learned period decomposition and applies inception-style 2D convolutions.
ConvTimeNet applies depthwise separable causal convolutions over multi-scale decomposed representations, with reversible instance normalization (RevIN) \cite{kim2022revin} on the input.
GPT4TS repurposes a frozen pretrained GPT-2 \cite{radford2019gpt2} backbone for time series forecasting by patching the input and mapping through the language model's intermediate layers.
TimeDiff and NsDiff are diffusion-based generative forecasting models that iteratively denoise future trajectories conditioned on the historical context.

\begin{table}[h]
\centering
\caption{Numerical forecasting model configurations. ``LR'' is the learning rate, ``Epochs'' is the maximum training budget, and ``Pat.'' is the early-stopping patience on validation MSE.}
\label{tab:numerical_models}
\small
\begin{tabular}{l l l r r r}
\toprule
Model & Type & Key Hyperparameters & LR & Epochs & Pat. \\
\midrule
Last Value & Naive & None & -- & -- & -- \\
Moving Average & Naive & window $=7$ & -- & -- & -- \\
DLinear & Linear & kernel $=\min(25,\,\max(3,\,2\lfloor L/2\rfloor + 1))$ & 1e-3 & 100 & 10 \\
PatchTST & Transformer & patch $=8$, stride $=4$, layers $=2$ & 1e-3 & 100 & 10 \\
iTransformer & Transformer & $d=64$, heads $=4$, layers $=2$, $d_\text{ff}=128$ & 1e-3 & 100 & 10 \\
TimesNet & CNN & $d=64$, $d_\text{ff}=64$, layers $=2$, top-$k=3$ & 1e-3 & 100 & 10 \\
ConvTimeNet & CNN & Official defaults, RevIN & 1e-4 & 10 & 3 \\
GPT4TS & LLM-based & Frozen GPT-2, patch $=4$, $d=768$ & 1e-4 & 10 & 5 \\
TimeDiff & Diffusion & Official defaults, point estimate $=$ median & 1e-4 & 10 & 5 \\
NsDiff & Diffusion & Official defaults, point estimate $=$ median & 1e-4 & 10 & 5 \\
\bottomrule
\end{tabular}
\end{table}

\subsection{Text-Augmented Forecasting Models}

Table~\ref{tab:text_augmented_models} summarizes the three text-augmented baselines and the CMA structure-aware reference baseline.
MM-TSF encodes each bin's text through BERT and pools the token-level outputs by averaging into an embedding, which is then integrated with the numerical backbone via linear interpolation controlled by a prompt-weight parameter.
GPT4MTS encodes each bin's text using BERT \texttt{[CLS]} embeddings and feeds them as soft prompts into a frozen GPT-2 backbone alongside patched time series tokens.
CAMEF applies RoBERTa with mean-pooling for text and MOMENT \cite{goswami2024moment} for series encoding, then stacks both as a two-token sequence processed by a GPT-2 fusion module.
All three baselines are evaluated under three input configurations, namely no text, flat text, and structured text, to expose how each architecture responds to textual content and to reply-chain structure within its own scale.

\begin{table}[h]
\centering
\caption{Text-augmented model configurations. ``LR'' denotes the learning rate, ``Epochs'' the maximum training epochs, and ``Patience'' the early-stopping patience on validation MSE.}
\label{tab:text_augmented_models}
\small
\begin{tabular}{lllrrr}
\toprule
Model & Text Encoder & Fusion Method & LR & Epochs & Patience \\
\midrule
MM-TSF & BERT (avg-pool) & Linear interpolation & 0.001 & 100 & 10 \\
GPT4MTS & BERT \texttt{[CLS]} & Prompt-based GPT-2 & 0.001 & 50 & 10 \\
CAMEF & RoBERTa (mean-pool) & GPT-2 two-token fusion & 0.0005 & 50 & 10 \\
CMA (ref.) & BERT \texttt{[CLS]} & Cross-modal attention & $1\!\times\!10^{-4}$ / $1\!\times\!10^{-3}$ & 100 & 10 \\
\bottomrule
\end{tabular}
\end{table}

\subsection{Shared Training Configuration}
Table~\ref{tab:shared_config} lists the training settings shared across all learned models.
Each experiment is repeated over five random seeds and results are averaged. Per-cell across-seed standard deviations for the numerical and text-augmented tracks are reported in Appendix~\ref{sec:appendix_seed_std}.
The data for each event is split chronologically into 70\% training, 10\% validation, and 20\% test segments. All learned models are optimized with Adam \cite{kingma2015adam}.

\begin{table}[h]
\centering
\caption{Shared experimental settings across granularities.}
\label{tab:shared_config}
\small
\begin{tabular}{lccc}
\toprule
& 1-Day & 12-Hour & 6-Hour \\
\midrule
Input length ($L$) & 14 & 28 & 56 \\
Prediction horizon ($H$) & 7 & 14 & 28 \\
Label length (decoder) & 7 & 14 & 28 \\
Batch size & 32 & 32 & 32 \\
Data split & \multicolumn{3}{c}{70\% / 10\% / 20\% (train / val / test)} \\
Loss / validation metric & \multicolumn{3}{c}{MSE} \\
Optimizer & \multicolumn{3}{c}{Adam} \\
\bottomrule
\end{tabular}
\end{table}

% 附录 E：CMA 模型细节
%
\section{CMA Model Details}
\label{sec:appendix_cma}

\begin{figure}[h]
\centering
\includegraphics[width=0.92\textwidth]{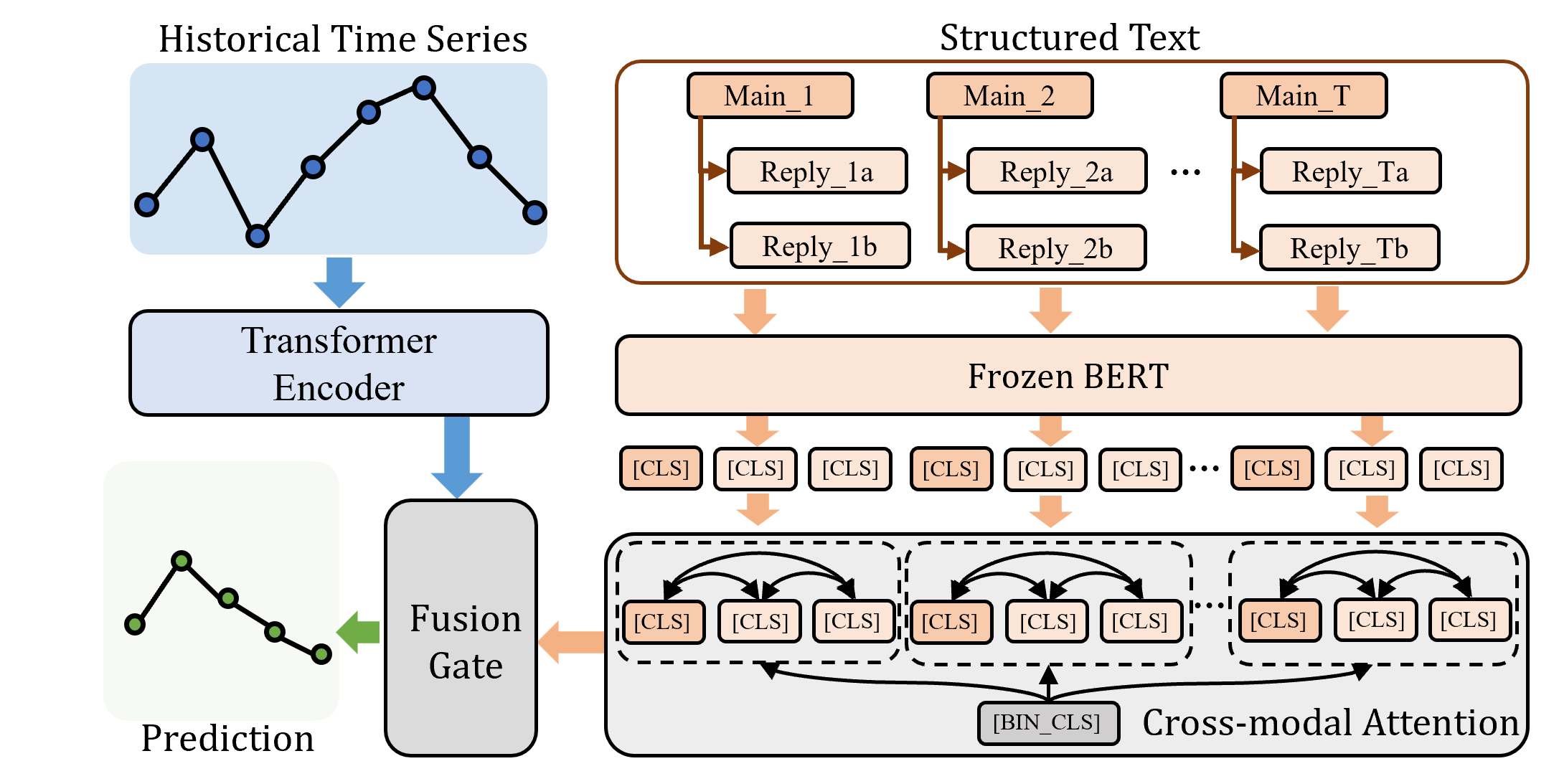}
\caption{CMA architecture overview. The numerical history is processed by a Transformer encoder (the standard decoder is bypassed). For each lookback bin, the per-post BERT \texttt{[CLS]} tokens are augmented with learned type embeddings (main post versus reply) and thread embeddings (which main-post thread the token belongs to within the bin), passed through an intra-bin self-attention layer, and pooled by a learnable \texttt{[BIN\_CLS]} attention query into a single bin vector. The per-bin vectors are added to the numerical encoder output at the matching lookback positions through a learnable per-position gate initialized at zero. The fused encoder output is then mapped to the horizon by a two-stage MLP, and a rolling-average prior is added as a residual. An auxiliary text-only head provides a side prediction from the bin vectors during training.}
\label{fig:cma_arch}
\end{figure}

This section gives the full architectural specification and training configuration of the CMA reference model introduced in Section~\ref{sec:exp_setup}.
CMA takes as input a numerical history $\mathbf{x}_{1:L}$ together with the bin-aligned text token sets $\{T_t\}_{t=1}^{L}$ described in Appendix~\ref{sec:appendix_ts_construction}, and produces a horizon prediction $\hat{\mathbf{y}}_{L+1:L+H}$.
The forward pass composes a Transformer encoder time-series backbone, a two-stage cross-modal pathway that summarizes each lookback bin into a structure-aware vector and fuses it with the numerical encoder output at the corresponding lookback position, and an encoder-to-horizon projection with a rolling-average prior residual:
\begin{equation}
\hat{\mathbf{y}}_{L+1:L+H}
\;=\;
\text{MLP}\!\left(\text{Fuse}\!\left(\text{TSEnc}(\mathbf{x}_{1:L}),\; \text{TXTEnc}(\{T_t\}_{t=1}^{L})\right)\right) + \mathbf{p}_{L+1:L+H},
\label{eq:cma_forward}
\end{equation}
where $\mathbf{p}_{L+1:L+H}$ is the rolling-average prior, and the overall architecture is illustrated in Figure~\ref{fig:cma_arch}.
CMA differs from the multimodal baselines MM-TSF, GPT4MTS, and CAMEF in two ways: (i) within each lookback bin, intra-bin self-attention with type and thread embeddings produces a structure-aware bin vector rather than a content-only mean of post embeddings, and (ii) the bin vector is fused with the numerical encoder output at the matching lookback position through a learnable per-position gate, so each lookback step decides for itself how much textual evidence to admit.

% Figure placeholder — actual diagram to be drawn later

\paragraph{Time-series backbone.}
The numerical history is processed by the encoder of a standard Transformer \cite{vaswani2017attention} encoder-decoder. The encoder embedding layer projects $\mathbf{x}_{1:L}$ together with its time features into $d_{\text{model}}$-dimensional representations, which are then refined by $e_{\text{layers}}$ encoder layers, producing $X_{\text{enc}} \in \mathbb{R}^{B \times L \times d_{\text{model}}}$. The standard Transformer decoder is not used in the CMA forward pass. The fused encoder output is mapped directly to the horizon by an MLP.

% ===== Per-Post Text Tokens =====
\subsection{Per-Post Text Tokens}

For each of the $L$ temporal bins in the lookback window, up to $K_{\text{post}} = 3$ main posts and up to $K_{\text{reply}} = 2$ replies per main post are selected from the posts that fall in the bin, yielding a maximum of $T_{\max} = K_{\text{post}}\,(1 + K_{\text{reply}}) = 9$ text tokens per bin. The selection criteria are deterministic rather than random: within each bin, main posts are ranked by their \emph{in-bin reply count} (the number of replies that also fall in the same bin) in descending order, and the top $K_{\text{post}}$ are kept; for each selected main post, the earliest $K_{\text{reply}}$ replies in chronological order are kept. The same selection rule is shared with the flat-text view used by the multimodal baselines (Appendix~\ref{sec:appendix_ts_construction}), so that all baselines see the same underlying post set and only differ in how that set is presented.
Each token is the BERT \texttt{[CLS]} embedding (768-dimensional) of one post or reply, produced by a frozen BERT encoder. The pooling is per-post: each post is summarized by its own \texttt{[CLS]} vector, and aggregation across posts within a bin is performed by the intra-bin encoder described in the next subsection. These token embeddings are precomputed and stored to avoid redundant encoding during training.
Within each bin, tokens are flattened in the order $[\text{main}_1, \text{reply}_{1a}, \text{reply}_{1b}, \text{main}_2, \ldots]$. Each token carries a binary type identifier $c_{t,m} \in \{0, 1\}$ (0 for main posts and 1 for replies, with repost tokens sharing the same identifier as replies) and a thread identifier $j_{t,m} \in \{0, \ldots, K_{\text{post}}-1\}$ that records which main-post thread the token belongs to within the bin (a main post and its replies share the same thread identifier). A per-token validity mask records padded positions in bins with fewer than $T_{\max}$ tokens.

\subsection{Two-Stage Cross-Modal Fusion}
\label{sec:appendix_cma_fusion}

The cross-modal pathway operates in two stages: an intra-bin encoder that summarizes each bin's text tokens into a single structure-aware vector, and a historical fusion module that injects these per-bin vectors into the numerical encoder output at the matching lookback positions.

\paragraph{Stage 1: intra-bin encoder.}
Stage 1 produces, for each of the $L$ lookback bins, a single $d_{\text{model}}$-dimensional bin vector that aggregates the bin's $T_{\max}$ post tokens while exposing reply-chain structure to the aggregation.
Token embeddings $E_{\text{text}} \in \mathbb{R}^{B \times L \times T_{\max} \times d_{\text{BERT}}}$ are first projected into the backbone space and combined with structural embeddings indexed by the per-token type and thread identifiers. Concretely, for the $m$-th token in bin $t$ with text embedding $\mathbf{e}_{t,m} \in \mathbb{R}^{d_{\text{BERT}}}$, type identifier $c_{t,m} \in \{0,1\}$, and thread identifier $j_{t,m} \in \{0, \ldots, K_{\text{post}}-1\}$:
\begin{equation}
\mathbf{h}_{t,m} \;=\; W_{\text{proj}}\, \mathbf{e}_{t,m} \;+\; \mathbf{E}^{\text{type}}_{c_{t,m}} \;+\; \mathbf{E}^{\text{thread}}_{j_{t,m}},
\end{equation}
where $W_{\text{proj}} \in \mathbb{R}^{d_{\text{model}} \times d_{\text{BERT}}}$ is a linear projection, $\mathbf{E}^{\text{type}} \in \mathbb{R}^{2 \times d_{\text{model}}}$ is a learned type-embedding table from which $\mathbf{E}^{\text{type}}_{c_{t,m}} \in \mathbb{R}^{d_{\text{model}}}$ is the row indexed by $c_{t,m}$, and $\mathbf{E}^{\text{thread}} \in \mathbb{R}^{K_{\text{post}} \times d_{\text{model}}}$ is a learned thread-embedding table indexed analogously by $j_{t,m}$. Stacking the per-token vectors yields $H \in \mathbb{R}^{B \times L \times T_{\max} \times d_{\text{model}}}$, with bin slice $H_t \in \mathbb{R}^{B \times T_{\max} \times d_{\text{model}}}$.
For each bin $t$, a multi-head self-attention layer with key-padding mask refines the token representations \emph{along the within-bin token axis}, i.e., attention is taken over the $T_{\max}$-dimensional token position; the $B$ and $L$ axes are treated as the batch dimension, so tokens from different bins do not interact at this stage:
\begin{equation}
H'_t \;=\; \text{LayerNorm}\!\left(H_t \;+\; \text{MHA}_{\text{token}}(H_t,\, H_t,\, H_t)\right), \qquad t = 1, \ldots, L,
\end{equation}
where the subscript ``token'' indicates that the queries, keys, and values are indexed by the within-bin token position $m \in \{1, \ldots, T_{\max}\}$, so tokens belonging to the same thread or to the same role can attend to one another and the bin representation becomes structure-aware.
The bin is then pooled by a learnable \texttt{[BIN\_CLS]} attention query that cross-attends along the same within-bin token axis with the same padding mask:
\begin{equation}
\mathbf{b}_t^{\text{base}} \;=\; \text{MHA}_{\text{token}}(\mathbf{q}_{\text{cls}},\, H'_t,\, H'_t),
\end{equation}
where $\mathbf{q}_{\text{cls}} \in \mathbb{R}^{d_{\text{model}}}$ is a single learned query that attends over the $T_{\max}$ tokens of bin $t$ to produce one $d_{\text{model}}$-dimensional vector per bin per sample.
On top of the base pool, a type-conditional residual pool encourages the bin vector to keep main-post and reply contributions explicitly separable. Main and reply tokens are mean-pooled separately into $\mathbf{b}_t^{\text{main}}$ and $\mathbf{b}_t^{\text{reply}}$, mixed by a linear layer, and added to the base pool through a learnable scalar $\alpha$ initialized at zero:
\begin{equation}
\mathbf{b}_t = \mathbf{b}_t^{\text{base}} + \alpha\,\bigl(W_{\text{mix}}[\mathbf{b}_t^{\text{main}};\,\mathbf{b}_t^{\text{reply}}] - \mathbf{b}_t^{\text{base}}\bigr),
\end{equation}
so that the module starts as the plain attention pool and learns whether to expand into the type-conditional form. The bin vector is zeroed for bins that contain no valid tokens, and a bin-validity mask $m_t \in \{0,1\}$ is recorded. Stacking over $t$ yields $B_{\text{text}} \in \mathbb{R}^{B \times L \times d_{\text{model}}}$ with mask $\mathbf{m} \in \{0,1\}^{B \times L}$.

\paragraph{Stage 2: historical fusion.}
Stage 2 injects the per-bin text vectors into the numerical encoder output at the matching lookback positions through a learnable per-position gate:
\begin{equation}
X_{\text{fused}}[:, t, :] \;=\; X_{\text{enc}}[:, t, :] \;+\; g_t \cdot m_t \cdot B_{\text{text}}[:, t, :], \qquad t = 1, \ldots, L,
\label{eq:cma_fusion}
\end{equation}
where $\mathbf{g} \in \mathbb{R}^{L}$ is a learnable scalar gate per lookback position, all initialized at zero so that the module is identity at the start of training. The per-position parameterization lets the model learn that, for example, recent bins or bins close to event onsets benefit more from textual evidence than quiescent bins.

\paragraph{Encoder-to-horizon projection.}
The fused encoder output $X_{\text{fused}} \in \mathbb{R}^{B \times L \times d_{\text{model}}}$ is mapped to the horizon prediction by a two-stage MLP that separates the temporal and feature dimensions:
\begin{align}
Z &= W_{\text{temp}}\, X_{\text{fused}}^\top, \\
\hat{\mathbf{y}}_{L+1:L+H} &= W_2\,\text{GELU}(W_1\, Z^\top) + \mathbf{p}_{L+1:L+H},
\end{align}
where $W_{\text{temp}} \in \mathbb{R}^{H \times L}$ projects along the temporal axis, $W_1 \in \mathbb{R}^{d_{\text{model}} \times d_{\text{model}}}$ and $W_2 \in \mathbb{R}^{c_{\text{out}} \times d_{\text{model}}}$ form a feature-wise MLP with dropout and GELU activation, and $\mathbf{p}$ is the rolling-average prior added as a residual.

% ===== Auxiliary Head and Text Modes =====
\subsection{Auxiliary Head and Input Configurations}

\paragraph{Text-only auxiliary head.}
To give the cross-modal pathway a direct supervised signal independent of the fusion gate, an auxiliary head predicts the target sequence from the bin vectors alone. The bin vectors are pooled along the lookback axis by a learned attention-weighted mean (with the bin-validity mask handling samples that have no text), and the pooled vector is mapped to $\hat{\mathbf{y}}^{\text{aux}} \in \mathbb{R}^{B \times H \times c_{\text{out}}}$ by a two-layer MLP with GELU activation. The auxiliary loss $\mathcal{L}_{\text{aux}} = \text{MSE}(\hat{\mathbf{y}}^{\text{aux}}, \mathbf{y}_{L+1:L+H})$ is added to the main MSE loss with weight $\lambda_{\text{aux}}$, so that the bin vectors are encouraged to be predictive of the target rather than only being instrumental to the fusion path:
\begin{equation}
\mathcal{L} = \text{MSE}(\hat{\mathbf{y}}, \mathbf{y}) + \lambda_{\text{aux}}\, \mathcal{L}_{\text{aux}}.
\end{equation}

% ===== Optimization and Hyperparameters =====
\subsection{Optimization and Hyperparameters}

\begin{table}[h]
\centering
\caption{CMA hyperparameters.}
\label{tab:cma_hyperparams}
\small
\begin{tabular}{lr}
\toprule
Parameter & Value \\
\midrule
Backbone $d_{\text{model}}$ & 512 \\
Backbone attention heads & 8 \\
Encoder layers ($e_{\text{layers}}$) & 2 \\
Feed-forward dimension ($d_{\text{ff}}$) & 2048 \\
Dropout & 0.1 \\
Intra-bin attention heads & 4 \\
Intra-bin self-attention layers & 1 \\
Text embedding dimension ($d_{\text{BERT}}$) & 768 \\
Max main posts per bin ($K_{\text{post}}$) & 3 \\
Max replies per main post ($K_{\text{reply}}$) & 2 \\
Max text tokens per bin ($T_{\max}$) & 9 \\
\midrule
Backbone learning rate & $1 \times 10^{-4}$ \\
Cross-modal block learning rate & $3 \times 10^{-4}$ \\
Auxiliary loss weight ($\lambda_{\text{aux}}$) & 0.05 \\
Max epochs & 100 \\
Early-stopping patience & 10 \\
Batch size & 32 \\
\bottomrule
\end{tabular}
\end{table}

\paragraph{Dual-optimizer schedule.}
CMA uses two separate Adam optimizers: one for the Transformer backbone parameters at a learning rate of $1 \times 10^{-4}$, and one for the cross-modal pathway, namely the intra-bin encoder, the historical fusion gate, the auxiliary head, and the encoder-to-horizon projection, at a higher learning rate of $3 \times 10^{-4}$. This separation allows the pretrained backbone components to be fine-tuned conservatively while the newly initialized cross-modal components converge faster. All gates and the type-pool residual coefficient $\alpha$ are initialized at zero so that the model starts at a no-text identity behavior and learns to inject text only where it helps.

% 附录 F：完整 benchmark 结果表 + Across-seed std
%
\section{Full Results}
\label{sec:appendix_results_tables}
\label{sec:appendix_seed_std}

This appendix lists the per-cell numerical tables behind the figure panels in Section~\ref{sec:exp_setup}, with each cell shown as the across-seed mean with the standard deviation in subscript. Tables~\ref{tab:numerical_results_std} and~\ref{tab:text_augmented_results_std} cover numerical and text-augmented forecasting (Figure~\ref{fig:num_text_results}); Tables~\ref{tab:mae_reply_all} and~\ref{tab:cross_event_all} cover the structure-aware reply-density MAE and the leave-one-category-out cross-event generalization (Figure~\ref{fig:mae_reply_cross_event}). All entries follow the unified per-event z-score scale described in Section~\ref{sec:exp_setup}.

\begin{table*}[t]
\centering
\caption{Per-cell across-seed mean and standard deviation for the numerical forecasting results. Each cell is shown as the mean over seeds with the standard deviation in subscript. Naive baselines are deterministic and have zero across-seed variance.}
\label{tab:numerical_results_std}
\resizebox{\linewidth}{!}{%
\begin{tabular}{l cc cc cc cc cc cc}
\toprule
\multirow{3}{*}{\textbf{Model}}
& \multicolumn{6}{c}{\textbf{\textit{Discussion Intensity}}}
& \multicolumn{6}{c}{\textbf{\textit{Sentiment Polarity}}} \\
\cmidrule(lr){2-7} \cmidrule(lr){8-13}
& \multicolumn{2}{c}{\textbf{1D}}
& \multicolumn{2}{c}{\textbf{12H}}
& \multicolumn{2}{c}{\textbf{6H}}
& \multicolumn{2}{c}{\textbf{1D}}
& \multicolumn{2}{c}{\textbf{12H}}
& \multicolumn{2}{c}{\textbf{6H}} \\
\cmidrule(lr){2-3} \cmidrule(lr){4-5} \cmidrule(lr){6-7}
\cmidrule(lr){8-9} \cmidrule(lr){10-11} \cmidrule(lr){12-13}
& MAE & MSE & MAE & MSE & MAE & MSE
& MAE & MSE & MAE & MSE & MAE & MSE \\
\midrule
Last Value & $0.115_{\pm 0.000}$ & $0.153_{\pm 0.000}$ & $0.076_{\pm 0.000}$ & $0.105_{\pm 0.000}$ & $0.068_{\pm 0.000}$ & $0.102_{\pm 0.000}$ & $0.465_{\pm 0.000}$ & $1.060_{\pm 0.000}$ & $0.403_{\pm 0.000}$ & $1.119_{\pm 0.000}$ & $0.379_{\pm 0.000}$ & $1.057_{\pm 0.000}$ \\
Moving Avg. & $0.129_{\pm 0.000}$ & $0.160_{\pm 0.000}$ & $0.077_{\pm 0.000}$ & $0.092_{\pm 0.000}$ & $0.064_{\pm 0.000}$ & $0.076_{\pm 0.000}$ & $0.492_{\pm 0.000}$ & $0.987_{\pm 0.000}$ & $0.423_{\pm 0.000}$ & $1.055_{\pm 0.000}$ & $0.373_{\pm 0.000}$ & $0.944_{\pm 0.000}$ \\
DLinear & $0.181_{\pm 0.012}$ & $0.128_{\pm 0.006}$ & $0.112_{\pm 0.008}$ & $0.083_{\pm 0.001}$ & $0.095_{\pm 0.002}$ & $0.071_{\pm 0.000}$ & $0.648_{\pm 0.007}$ & $0.888_{\pm 0.006}$ & $0.560_{\pm 0.006}$ & $0.932_{\pm 0.001}$ & $0.564_{\pm 0.013}$ & $0.888_{\pm 0.004}$ \\
PatchTST & $0.153_{\pm 0.008}$ & $0.126_{\pm 0.003}$ & $0.106_{\pm 0.009}$ & $0.085_{\pm 0.001}$ & $0.098_{\pm 0.002}$ & $0.072_{\pm 0.001}$ & $0.591_{\pm 0.008}$ & $0.849_{\pm 0.004}$ & $0.528_{\pm 0.004}$ & $0.983_{\pm 0.042}$ & $0.530_{\pm 0.035}$ & $0.941_{\pm 0.030}$ \\
iTransformer & $0.156_{\pm 0.014}$ & $0.124_{\pm 0.002}$ & $0.099_{\pm 0.001}$ & $0.087_{\pm 0.000}$ & $0.090_{\pm 0.002}$ & $0.073_{\pm 0.001}$ & $0.585_{\pm 0.013}$ & $0.870_{\pm 0.010}$ & $0.539_{\pm 0.019}$ & $0.977_{\pm 0.032}$ & $0.544_{\pm 0.036}$ & $0.938_{\pm 0.017}$ \\
TimesNet & $0.155_{\pm 0.011}$ & $0.122_{\pm 0.003}$ & $0.118_{\pm 0.019}$ & $0.097_{\pm 0.004}$ & $0.119_{\pm 0.010}$ & $0.085_{\pm 0.006}$ & $0.687_{\pm 0.058}$ & $0.941_{\pm 0.049}$ & $0.556_{\pm 0.045}$ & $0.975_{\pm 0.061}$ & $0.579_{\pm 0.039}$ & $1.022_{\pm 0.102}$ \\
ConvTimeNet & $0.194_{\pm 0.002}$ & $0.126_{\pm 0.001}$ & $0.133_{\pm 0.021}$ & $0.085_{\pm 0.003}$ & $0.116_{\pm 0.011}$ & $0.074_{\pm 0.002}$ & $0.672_{\pm 0.015}$ & $0.899_{\pm 0.015}$ & $0.553_{\pm 0.042}$ & $0.936_{\pm 0.011}$ & $0.580_{\pm 0.031}$ & $0.892_{\pm 0.026}$ \\
GPT4TS & $0.117_{\pm 0.002}$ & $0.131_{\pm 0.003}$ & $0.074_{\pm 0.003}$ & $0.083_{\pm 0.002}$ & $0.059_{\pm 0.001}$ & $0.068_{\pm 0.000}$ & $0.497_{\pm 0.004}$ & $0.904_{\pm 0.004}$ & $0.428_{\pm 0.002}$ & $0.937_{\pm 0.004}$ & $0.398_{\pm 0.002}$ & $0.859_{\pm 0.003}$ \\
TimeDiff & $0.150_{\pm 0.006}$ & $0.129_{\pm 0.005}$ & $0.090_{\pm 0.002}$ & $0.085_{\pm 0.001}$ & $0.111_{\pm 0.011}$ & $0.075_{\pm 0.001}$ & $0.638_{\pm 0.045}$ & $1.003_{\pm 0.200}$ & $0.601_{\pm 0.040}$ & $1.002_{\pm 0.200}$ & $0.573_{\pm 0.040}$ & $0.908_{\pm 0.180}$ \\
NsDiff & $0.157_{\pm 0.003}$ & $0.131_{\pm 0.007}$ & $0.094_{\pm 0.002}$ & $0.089_{\pm 0.008}$ & $0.112_{\pm 0.003}$ & $0.079_{\pm 0.011}$ & $0.565_{\pm 0.007}$ & $0.875_{\pm 0.005}$ & $0.558_{\pm 0.012}$ & $0.984_{\pm 0.004}$ & $0.547_{\pm 0.007}$ & $0.962_{\pm 0.015}$ \\
\bottomrule
\end{tabular}%
}
\end{table*}

\begin{table*}[t]
\centering
\caption{Per-cell across-seed mean and standard deviation for the text-augmented forecasting results. Each cell is shown as the mean over seeds with the standard deviation in subscript.}
\label{tab:text_augmented_results_std}
\small
\setlength{\tabcolsep}{6pt}
\begin{tabular}{ll cc}
\toprule
\textbf{Model} & \textbf{Text}
& \textbf{\textit{Discussion Intensity}}
& \textbf{\textit{Sentiment Polarity}} \\
\midrule
\multirow{3}{*}{MM-TSF}
& No Text     & $0.454_{\pm 0.008}$ & $0.552_{\pm 0.006}$ \\
& Flat        & $0.446_{\pm 0.006}$ & $0.536_{\pm 0.004}$ \\
& Structured  & $0.445_{\pm 0.005}$ & $0.508_{\pm 0.006}$ \\
\midrule
\multirow{3}{*}{GPT4MTS}
& No Text     & $0.238_{\pm 0.001}$ & $0.611_{\pm 0.004}$ \\
& Flat        & $0.236_{\pm 0.000}$ & $0.610_{\pm 0.002}$ \\
& Structured  & $0.236_{\pm 0.001}$ & $0.607_{\pm 0.004}$ \\
\midrule
\multirow{3}{*}{CAMEF}
& No Text     & $0.275_{\pm 0.011}$ & $0.686_{\pm 0.024}$ \\
& Flat        & $0.273_{\pm 0.005}$ & $0.675_{\pm 0.011}$ \\
& Structured  & $0.271_{\pm 0.014}$ & $0.664_{\pm 0.011}$ \\
\midrule
\multirow{3}{*}{CMA}
& No Text     & $0.241_{\pm 0.003}$ & $0.297_{\pm 0.003}$ \\
& Flat        & $0.240_{\pm 0.002}$ & $0.275_{\pm 0.002}$ \\
& Structured  & $0.232_{\pm 0.003}$ & $0.266_{\pm 0.003}$ \\
\bottomrule
\end{tabular}
\end{table*}

\begin{table*}[t]
\centering
\caption{Structure-aware reply-density MAE on Sentiment Polarity at 1-day granularity.}
\label{tab:mae_reply_all}
\small
\setlength{\tabcolsep}{6pt}
\begin{tabular}{l cccc}
\toprule
\textbf{Model} & \textbf{5\%} & \textbf{10\%} & \textbf{20\%} & \textbf{50\%} \\
\midrule
CMA\textsubscript{no\_text}    & $0.587_{\pm 0.009}$ & $0.505_{\pm 0.006}$ & $0.412_{\pm 0.005}$ & $0.355_{\pm 0.004}$ \\
CMA\textsubscript{flat}        & $0.587_{\pm 0.009}$ & $0.503_{\pm 0.006}$ & $0.410_{\pm 0.005}$ & $0.355_{\pm 0.004}$ \\
CMA\textsubscript{struct}      & $\mathbf{0.588}_{\pm 0.009}$ & $\mathbf{0.502}_{\pm 0.006}$ & $\mathbf{0.409}_{\pm 0.005}$ & $\mathbf{0.353}_{\pm 0.004}$ \\
\midrule
Last Value     & $1.204_{\pm 0.000}$ & $0.995_{\pm 0.000}$ & $0.757_{\pm 0.000}$ & $0.637_{\pm 0.000}$ \\
Moving Average & $1.259_{\pm 0.000}$ & $0.991_{\pm 0.000}$ & $0.720_{\pm 0.000}$ & $0.656_{\pm 0.000}$ \\
DLinear        & $1.130_{\pm 0.025}$ & $0.920_{\pm 0.017}$ & $0.664_{\pm 0.011}$ & $0.587_{\pm 0.009}$ \\
GPT4TS         & $1.197_{\pm 0.018}$ & $0.944_{\pm 0.011}$ & $0.684_{\pm 0.008}$ & $0.597_{\pm 0.006}$ \\
\midrule
MM-TSF         & $1.056_{\pm 0.024}$ & $0.853_{\pm 0.015}$ & $0.637_{\pm 0.011}$ & $0.526_{\pm 0.008}$ \\
GPT4MTS        & $1.442_{\pm 0.015}$ & $1.180_{\pm 0.010}$ & $0.846_{\pm 0.007}$ & $0.763_{\pm 0.005}$ \\
CAMEF          & $1.357_{\pm 0.041}$ & $1.100_{\pm 0.026}$ & $0.786_{\pm 0.017}$ & $0.721_{\pm 0.014}$ \\
\bottomrule
\end{tabular}
\end{table*}

\begin{table*}[t]
\centering
\caption{Leave-one-category-out cross-event generalization at 1-day granularity. Each cell reports MAE on the per-event z-score scale. Columns Nat., Pol., Soc., Tech., Sports, and Avg. denote Natural Disaster, Political, Social Movement, Technology, Sports \& Entertainment, and the category average, respectively. Best results in \textbf{bold}, second best \underline{underlined}.}
\label{tab:cross_event_all}
\small
\resizebox{\linewidth}{!}{%
\begin{tabular}{ll cccccc}
\toprule
\textbf{Target} & \textbf{Model}
& \textbf{Nat.} & \textbf{Pol.} & \textbf{Soc.} & \textbf{Tech.} & \textbf{Sports} & \textbf{Avg.} \\
\midrule
\multirow{8}{*}{\textit{Discussion Intensity}}
& DLinear      & $\mathbf{0.289}_{\pm 0.004}$ & $\underline{0.345}_{\pm 0.005}$ & $\mathbf{0.190}_{\pm 0.003}$ & $\mathbf{0.245}_{\pm 0.004}$ & $\underline{0.291}_{\pm 0.004}$ & $\underline{0.272}_{\pm 0.004}$ \\
& PatchTST     & $\underline{0.257}_{\pm 0.004}$ & $\mathbf{0.331}_{\pm 0.005}$ & $\underline{0.257}_{\pm 0.004}$ & $\underline{0.275}_{\pm 0.004}$ & $\mathbf{0.213}_{\pm 0.003}$ & $\mathbf{0.266}_{\pm 0.004}$ \\
& TimesNet     & $0.260_{\pm 0.018}$ & $0.363_{\pm 0.025}$ & $0.267_{\pm 0.019}$ & $0.310_{\pm 0.022}$ & $0.256_{\pm 0.018}$ & $0.291_{\pm 0.020}$ \\
& GPT4TS       & $0.370_{\pm 0.004}$ & $0.498_{\pm 0.005}$ & $0.303_{\pm 0.003}$ & $0.371_{\pm 0.004}$ & $0.304_{\pm 0.003}$ & $0.369_{\pm 0.004}$ \\
& MM-TSF       & $0.414_{\pm 0.006}$ & $0.483_{\pm 0.007}$ & $0.330_{\pm 0.005}$ & $0.416_{\pm 0.006}$ & $0.324_{\pm 0.005}$ & $0.394_{\pm 0.006}$ \\
& GPT4MTS      & $0.295_{\pm 0.002}$ & $0.384_{\pm 0.003}$ & $0.242_{\pm 0.002}$ & $0.285_{\pm 0.002}$ & $0.233_{\pm 0.002}$ & $0.288_{\pm 0.002}$ \\
& CAMEF        & $0.413_{\pm 0.008}$ & $0.419_{\pm 0.008}$ & $0.303_{\pm 0.006}$ & $0.312_{\pm 0.006}$ & $0.334_{\pm 0.007}$ & $0.356_{\pm 0.007}$ \\
& CMA          & $0.255_{\pm 0.003}$ & $0.392_{\pm 0.004}$ & $0.247_{\pm 0.002}$ & $0.319_{\pm 0.003}$ & $0.226_{\pm 0.002}$ & $0.288_{\pm 0.003}$ \\
\midrule
\multirow{8}{*}{\textit{Sentiment Polarity}}
& DLinear      & $0.887_{\pm 0.013}$ & $\underline{0.714}_{\pm 0.011}$ & $0.701_{\pm 0.011}$ & $\underline{0.597}_{\pm 0.009}$ & $\underline{0.588}_{\pm 0.009}$ & $0.697_{\pm 0.010}$ \\
& PatchTST     & $0.882_{\pm 0.013}$ & $\underline{0.705}_{\pm 0.011}$ & $0.619_{\pm 0.009}$ & $0.711_{\pm 0.011}$ & $0.599_{\pm 0.009}$ & $0.703_{\pm 0.011}$ \\
& TimesNet     & $0.877_{\pm 0.061}$ & $0.711_{\pm 0.050}$ & $0.776_{\pm 0.054}$ & $0.715_{\pm 0.050}$ & $0.601_{\pm 0.042}$ & $0.736_{\pm 0.052}$ \\
& GPT4TS       & $0.709_{\pm 0.007}$ & $0.560_{\pm 0.006}$ & $0.474_{\pm 0.005}$ & $0.454_{\pm 0.005}$ & $0.409_{\pm 0.004}$ & $0.521_{\pm 0.005}$ \\
& MM-TSF       & $0.611_{\pm 0.009}$ & $0.533_{\pm 0.008}$ & $0.519_{\pm 0.008}$ & $0.439_{\pm 0.007}$ & $0.430_{\pm 0.006}$ & $0.506_{\pm 0.008}$ \\
& CAMEF        & $0.852_{\pm 0.017}$ & $0.739_{\pm 0.015}$ & $0.745_{\pm 0.015}$ & $0.605_{\pm 0.012}$ & $0.633_{\pm 0.013}$ & $0.715_{\pm 0.014}$ \\
& GPT4MTS      & $0.871_{\pm 0.006}$ & $0.729_{\pm 0.005}$ & $0.553_{\pm 0.004}$ & $0.577_{\pm 0.004}$ & $0.487_{\pm 0.003}$ & $0.643_{\pm 0.005}$ \\
& CMA          & $0.373_{\pm 0.004}$ & $\mathbf{0.344}_{\pm 0.003}$ & $\mathbf{0.244}_{\pm 0.002}$ & $\mathbf{0.266}_{\pm 0.003}$ & $0.240_{\pm 0.002}$ & $0.293_{\pm 0.003}$ \\
\bottomrule
\end{tabular}
}
\end{table*}

% 附录 G：Per-Event 可视化与文本示例
%
\section{Per-Event Visualizations and Text Examples}
\label{sec:appendix_visualizations}

This appendix presents multi-granularity visualizations for 15 representative events (three per category).
Each figure shows six subplots arranged as three rows (1-Day, 12-Hour, 6-Hour) and two columns: the left column plots Discussion Intensity (post volume) and the right column plots Sentiment Polarity. Empty bins are forward-filled within each split independently in the benchmark pipeline (Appendix~\ref{sec:appendix_ts_construction}).

\begin{figure}[b]
\centering
\includegraphics[width=\textwidth]{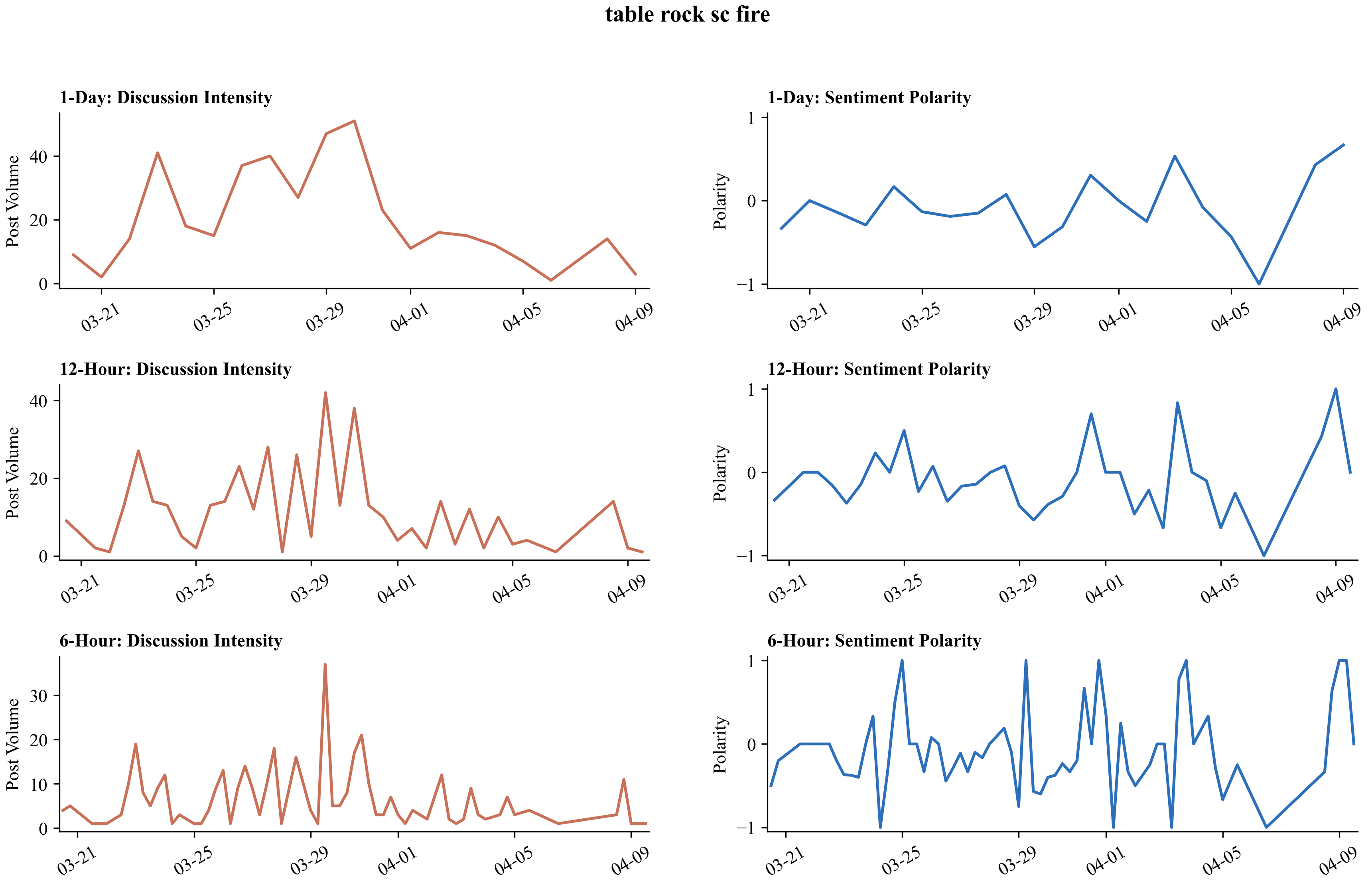}
\caption{Discussion Intensity and Sentiment Polarity at 1-Day, 12-Hour, and 6-Hour granularities for Table Rock Fire, South Carolina.}
\label{fig:appendix_table_rock_fire}
\end{figure}

\begin{figure}
\centering
\includegraphics[width=\textwidth]{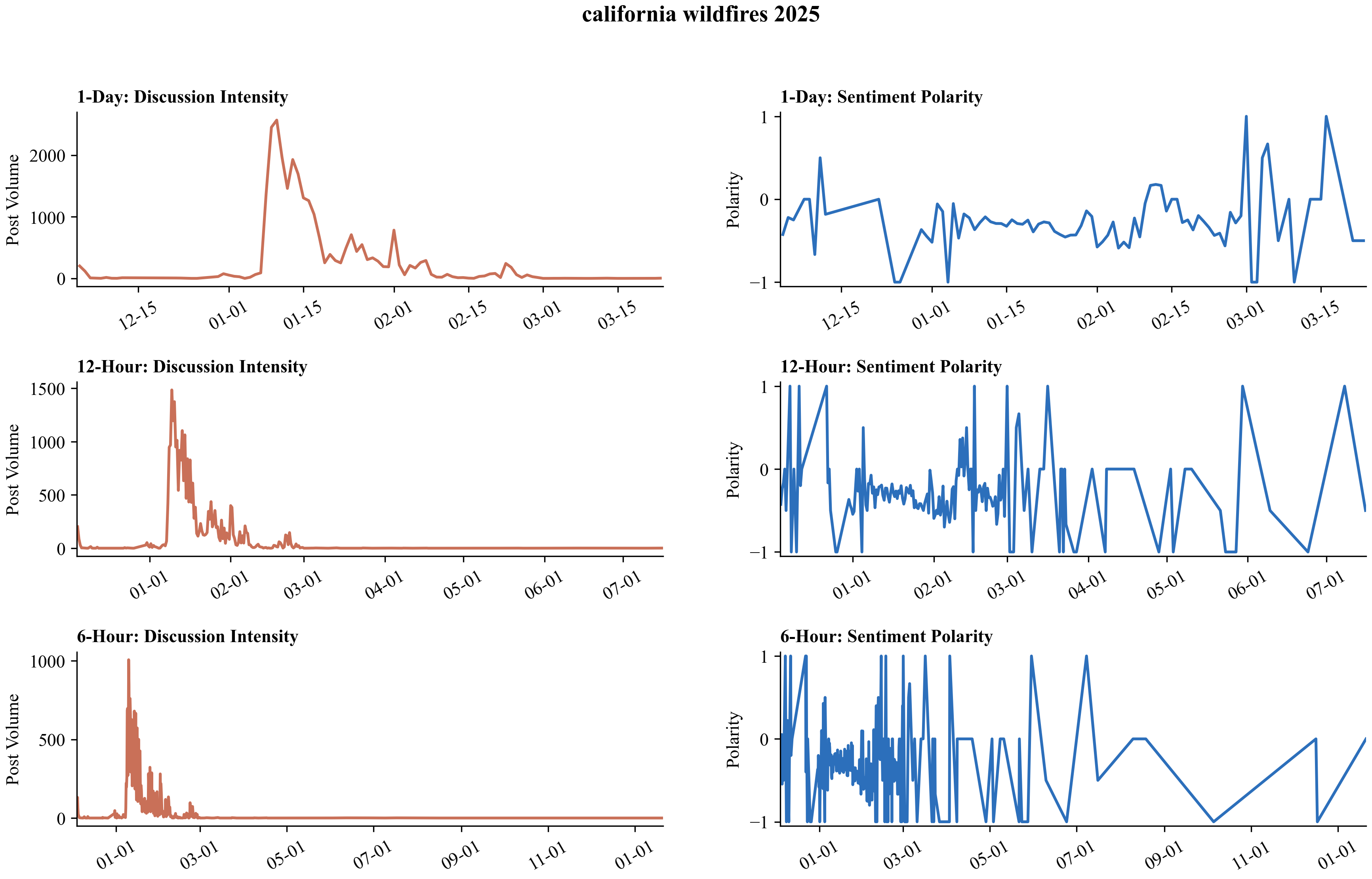}
\caption{Discussion Intensity and Sentiment Polarity at 1-Day, 12-Hour, and 6-Hour granularities for California Wildfires 2025.}
\label{fig:appendix_california_wildfires}
\end{figure}

\begin{figure}
\centering
\includegraphics[width=\textwidth]{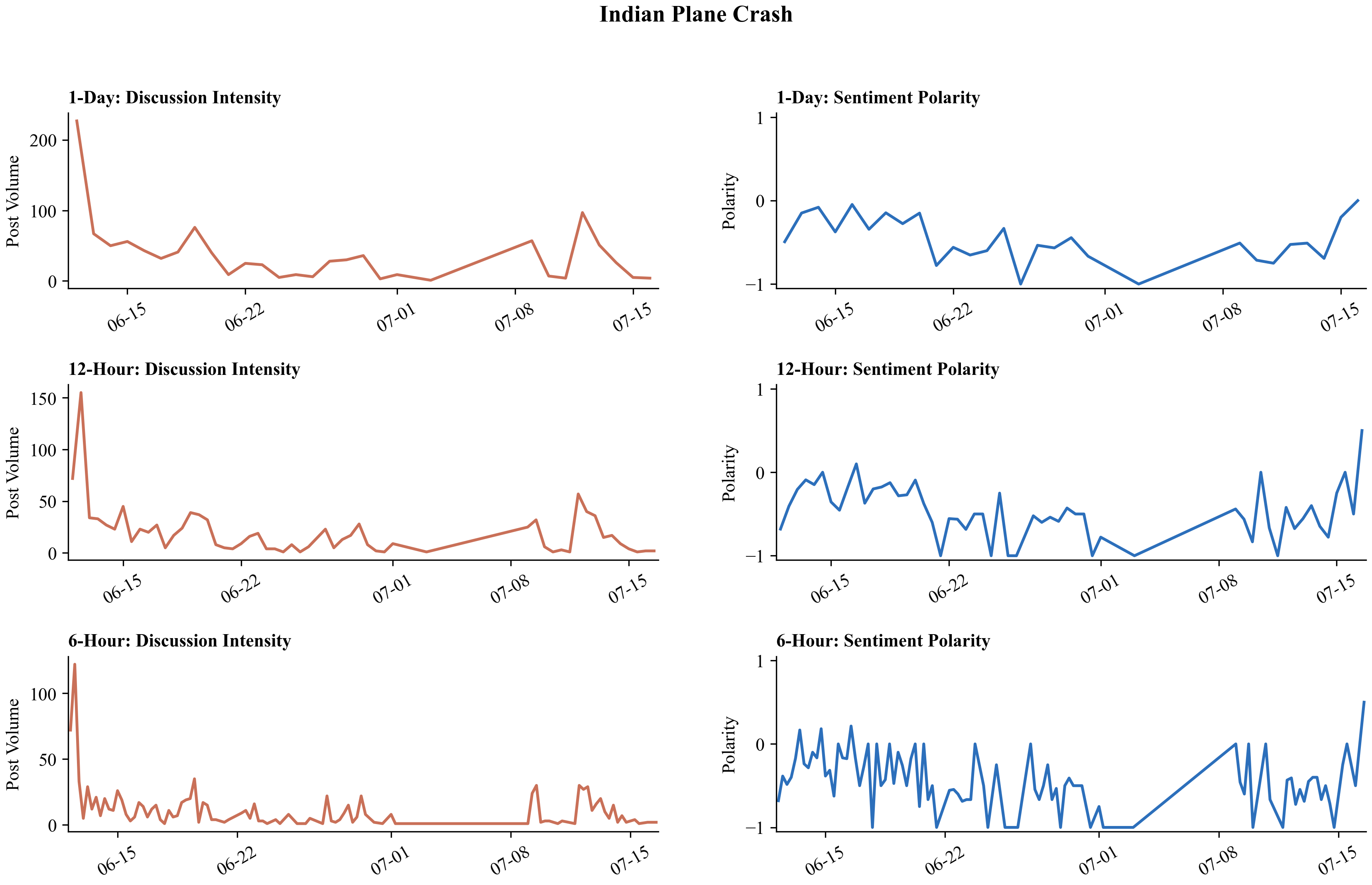}
\caption{Discussion Intensity and Sentiment Polarity at 1-Day, 12-Hour, and 6-Hour granularities for Indian Plane Crash.}
\label{fig:appendix_indian_plane_crash}
\end{figure}

\begin{figure}
\centering
\includegraphics[width=\textwidth]{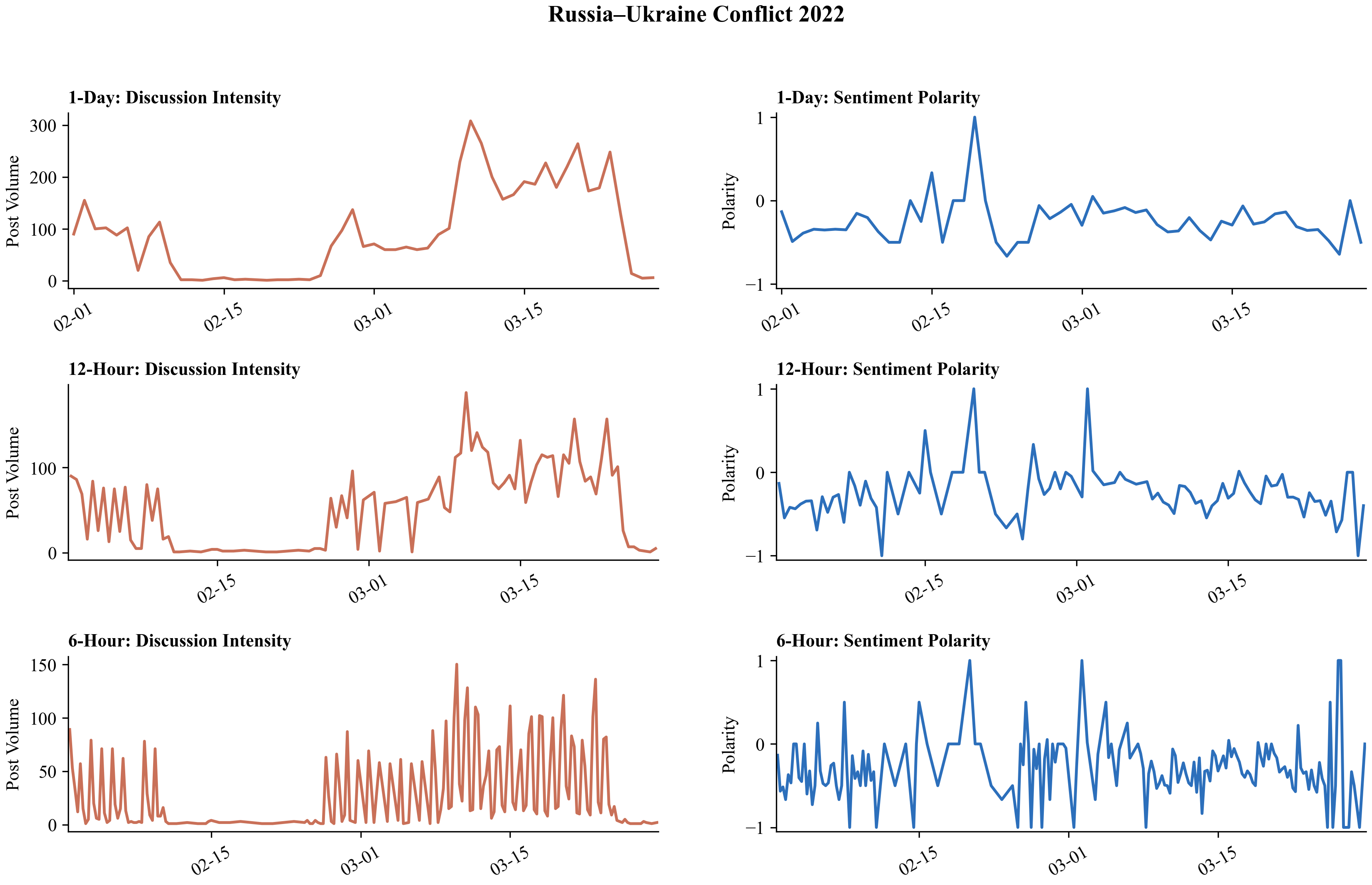}
\caption{Discussion Intensity and Sentiment Polarity at 1-Day, 12-Hour, and 6-Hour granularities for Russia--Ukraine Conflict 2022.}
\label{fig:appendix_ukraine_invasion}
\end{figure}

\begin{figure}
\centering
\includegraphics[width=\textwidth]{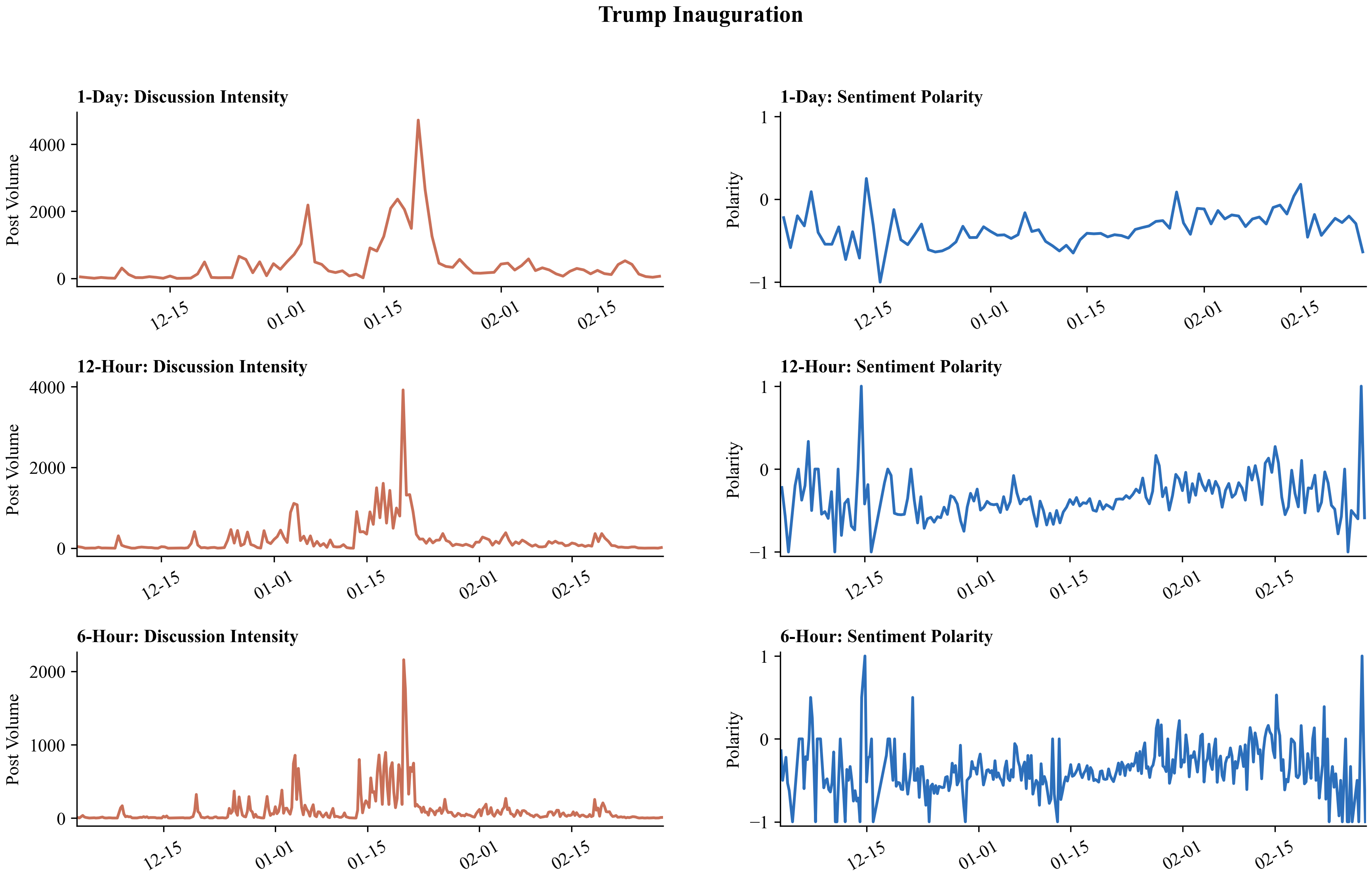}
\caption{Discussion Intensity and Sentiment Polarity at 1-Day, 12-Hour, and 6-Hour granularities for Trump Inauguration.}
\label{fig:appendix_trump_inauguration}
\end{figure}

\begin{figure}
\centering
\includegraphics[width=\textwidth]{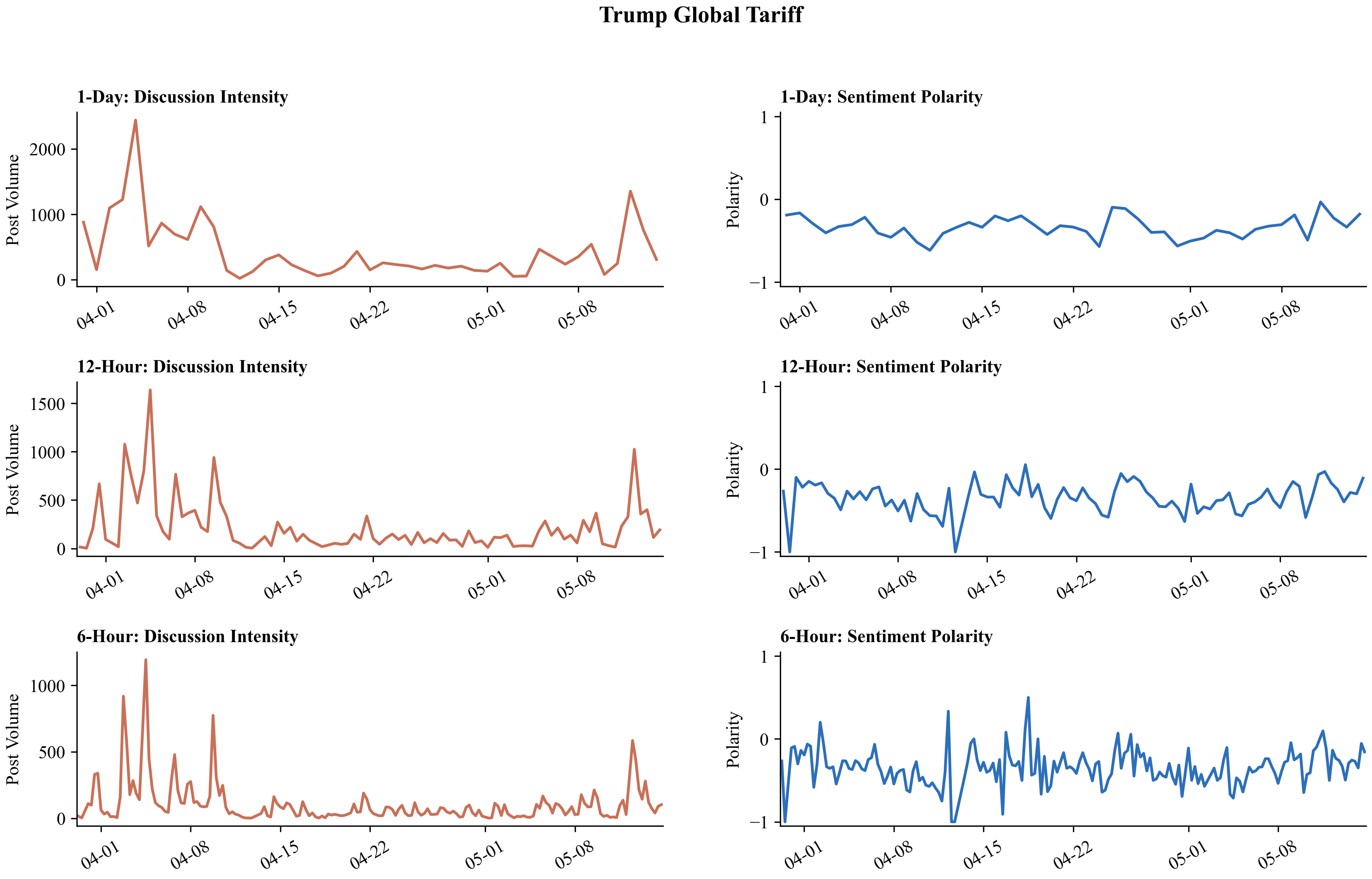}
\caption{Discussion Intensity and Sentiment Polarity at 1-Day, 12-Hour, and 6-Hour granularities for Trump Global Tariff.}
\label{fig:appendix_trump_tariff}
\end{figure}

\begin{figure}
\centering
\includegraphics[width=\textwidth]{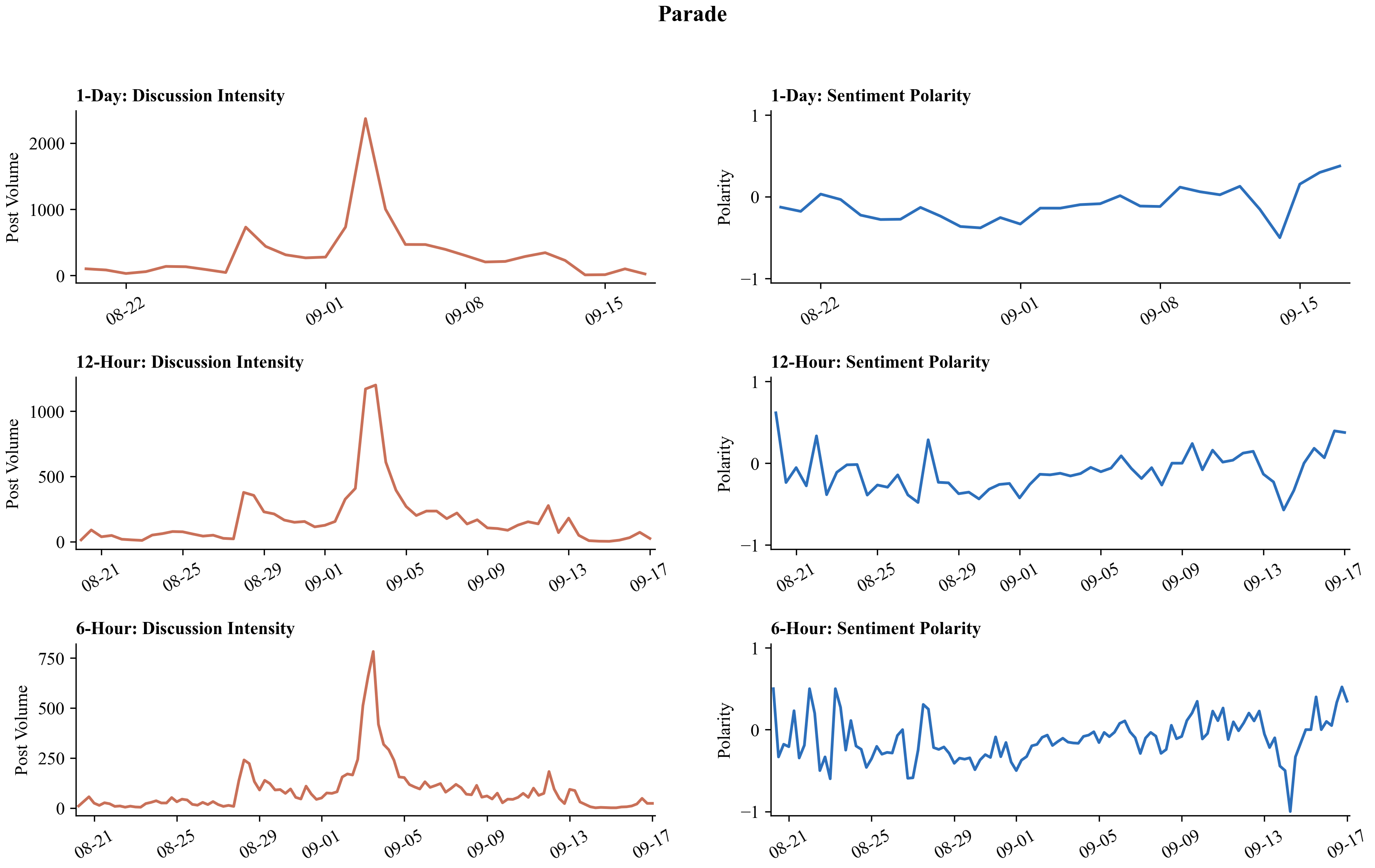}
\caption{Discussion Intensity and Sentiment Polarity at 1-Day, 12-Hour, and 6-Hour granularities for Parade.}
\label{fig:appendix_parade}
\end{figure}

\begin{figure}
\centering
\includegraphics[width=\textwidth]{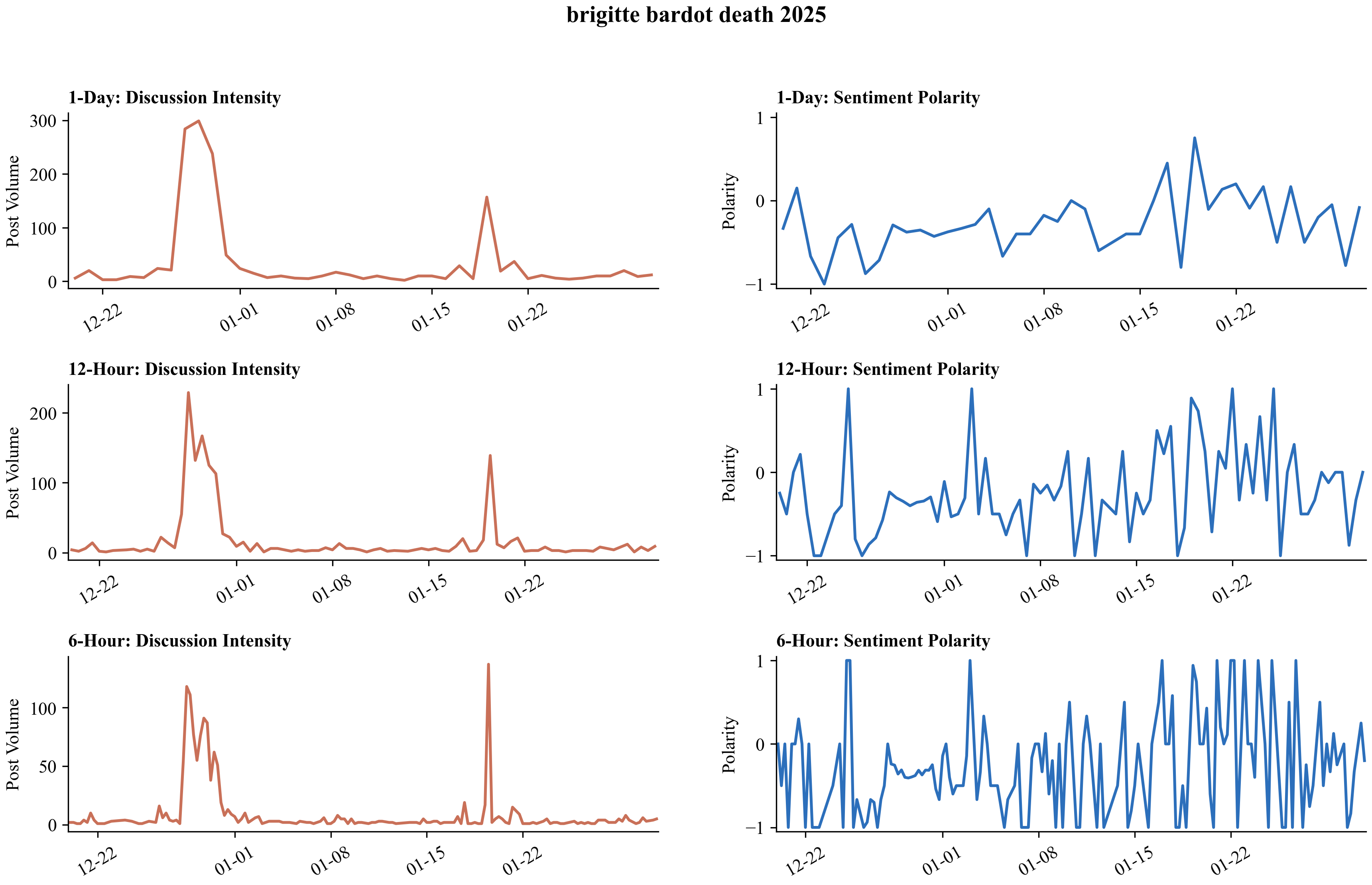}
\caption{Discussion Intensity and Sentiment Polarity at 1-Day, 12-Hour, and 6-Hour granularities for Brigitte Bardot Death 2025.}
\label{fig:appendix_brigitte_bardot}
\end{figure}

\begin{figure}
\centering
\includegraphics[width=\textwidth]{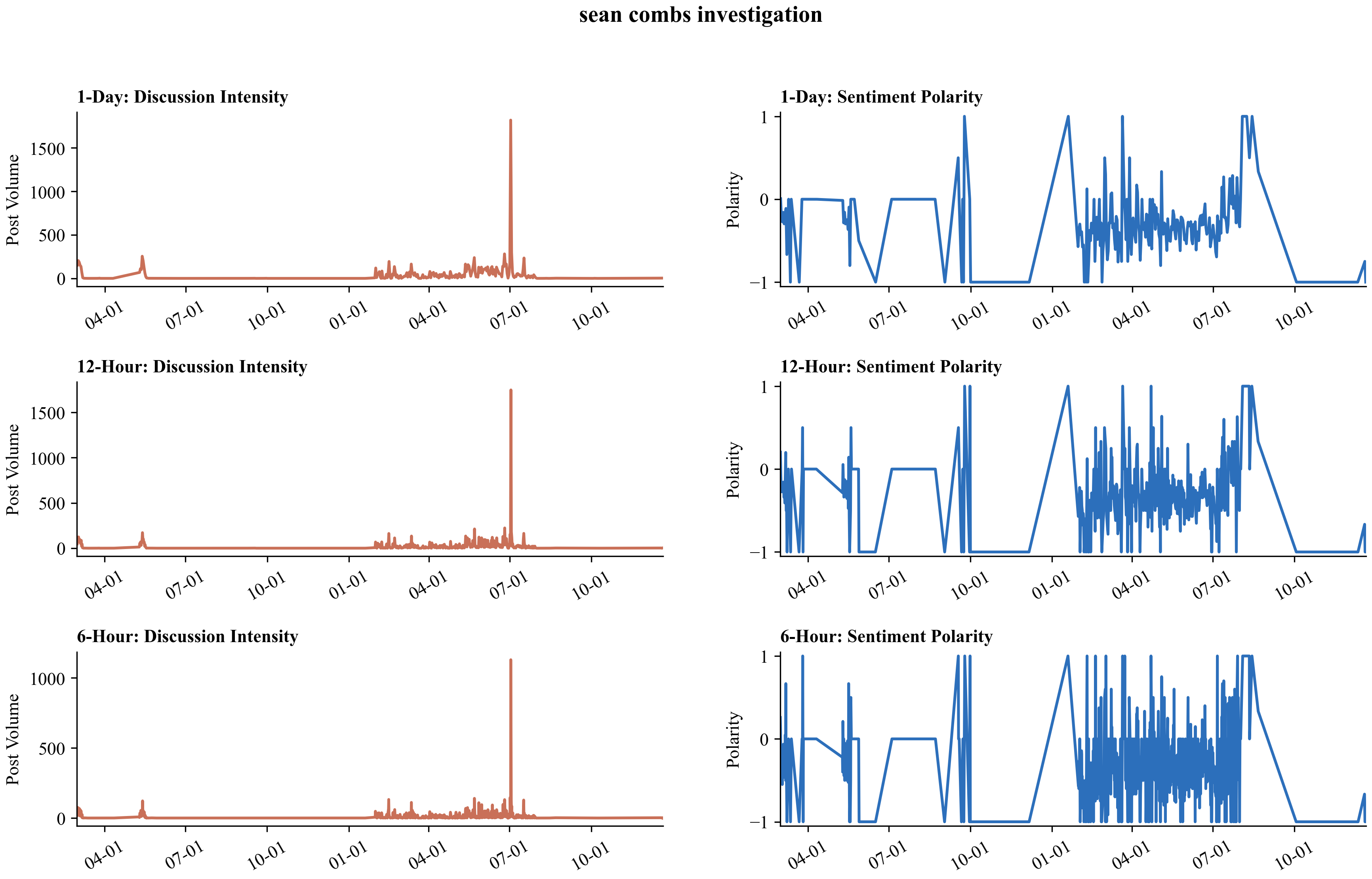}
\caption{Discussion Intensity and Sentiment Polarity at 1-Day, 12-Hour, and 6-Hour granularities for Sean Combs Investigation.}
\label{fig:appendix_sean_combs}
\end{figure}

\begin{figure}
\centering
\includegraphics[width=\textwidth]{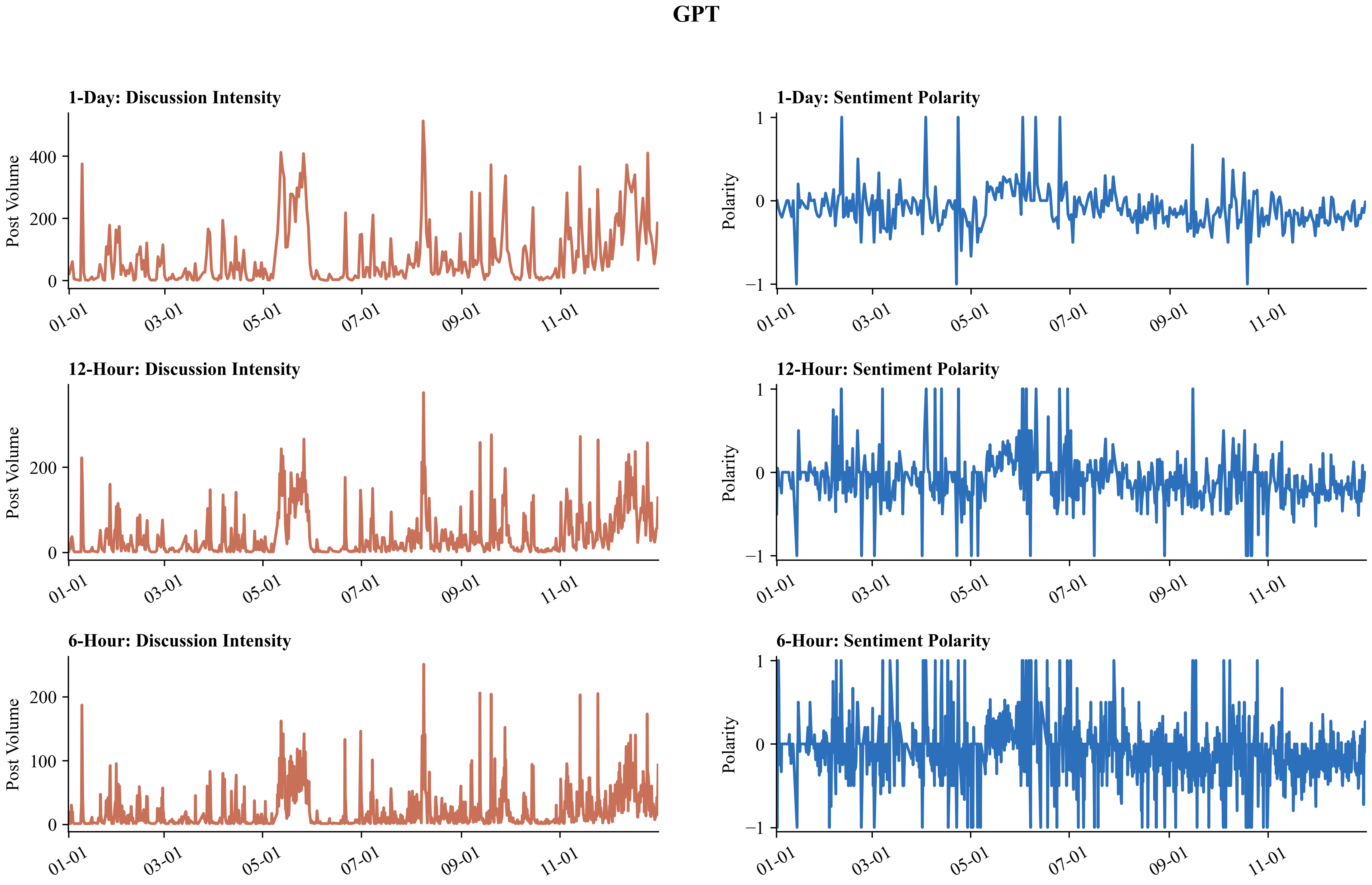}
\caption{Discussion Intensity and Sentiment Polarity at 1-Day, 12-Hour, and 6-Hour granularities for GPT.}
\label{fig:appendix_gpt}
\end{figure}

\begin{figure}
\centering
\includegraphics[width=\textwidth]{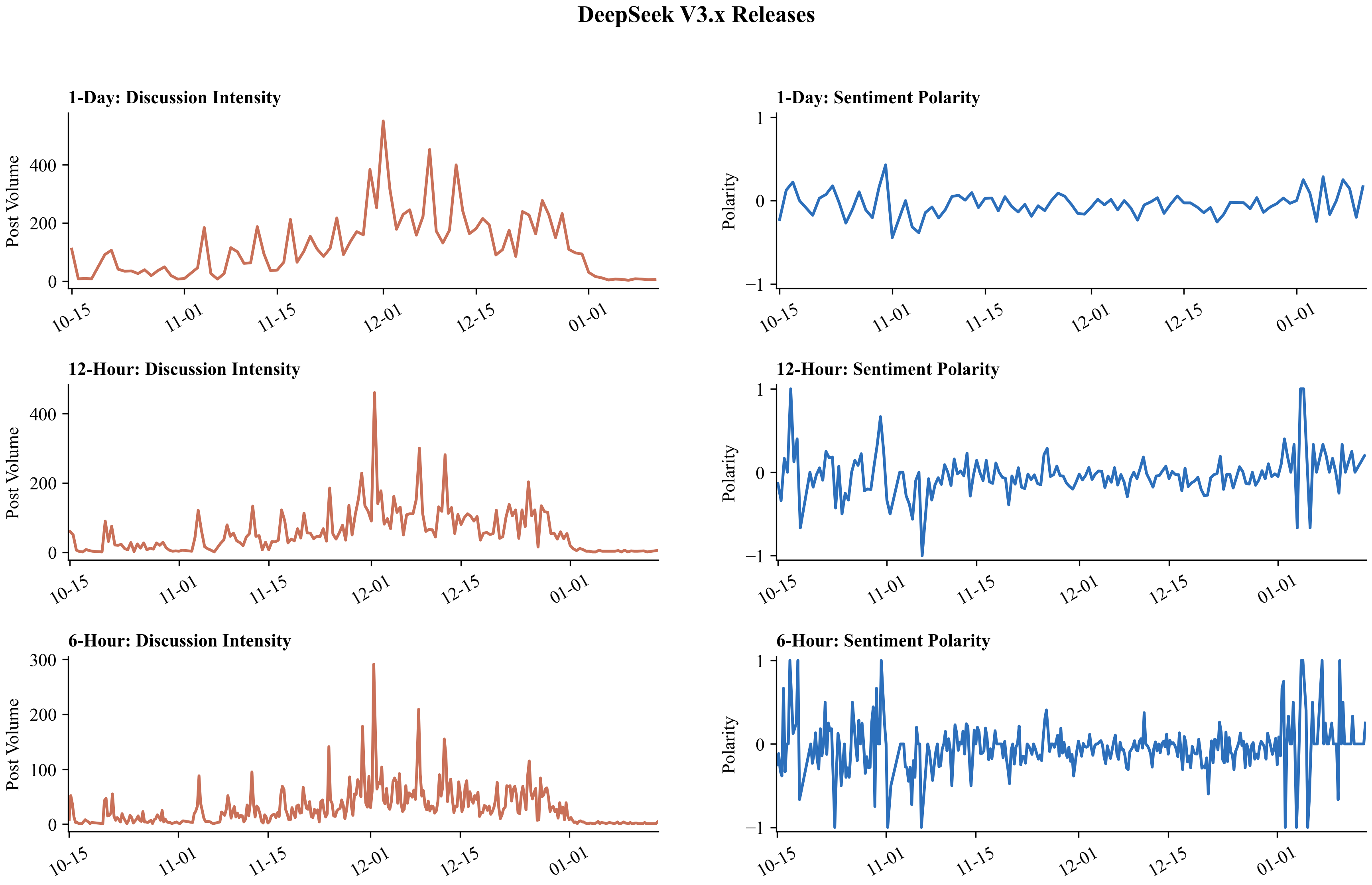}
\caption{Discussion Intensity and Sentiment Polarity at 1-Day, 12-Hour, and 6-Hour granularities for DeepSeek V3.x Releases.}
\label{fig:appendix_deepseek}
\end{figure}

\begin{figure}
\centering
\includegraphics[width=\textwidth]{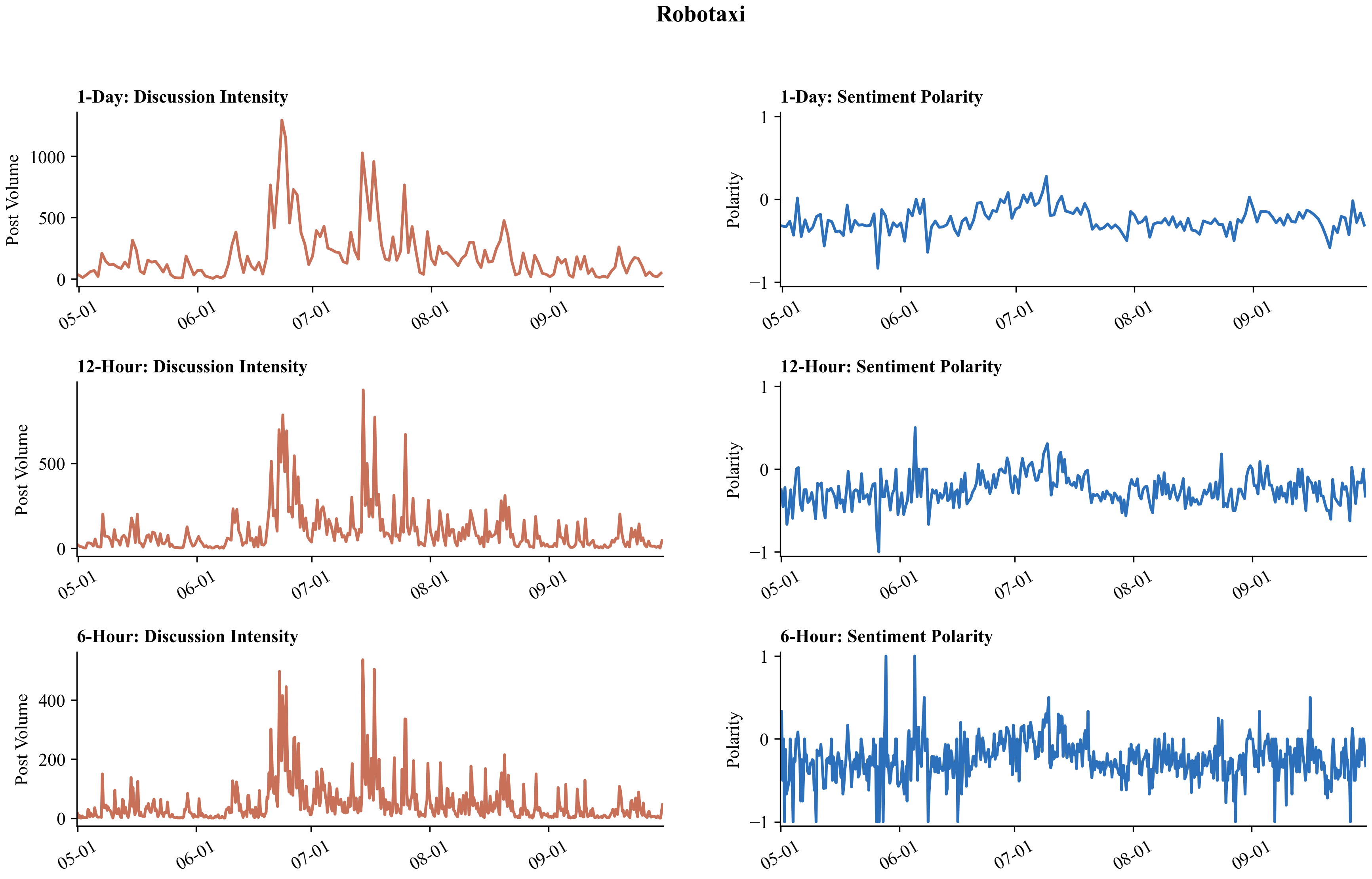}
\caption{Discussion Intensity and Sentiment Polarity at 1-Day, 12-Hour, and 6-Hour granularities for Robotaxi.}
\label{fig:appendix_robotaxi}
\end{figure}

\begin{figure}
\centering
\includegraphics[width=\textwidth]{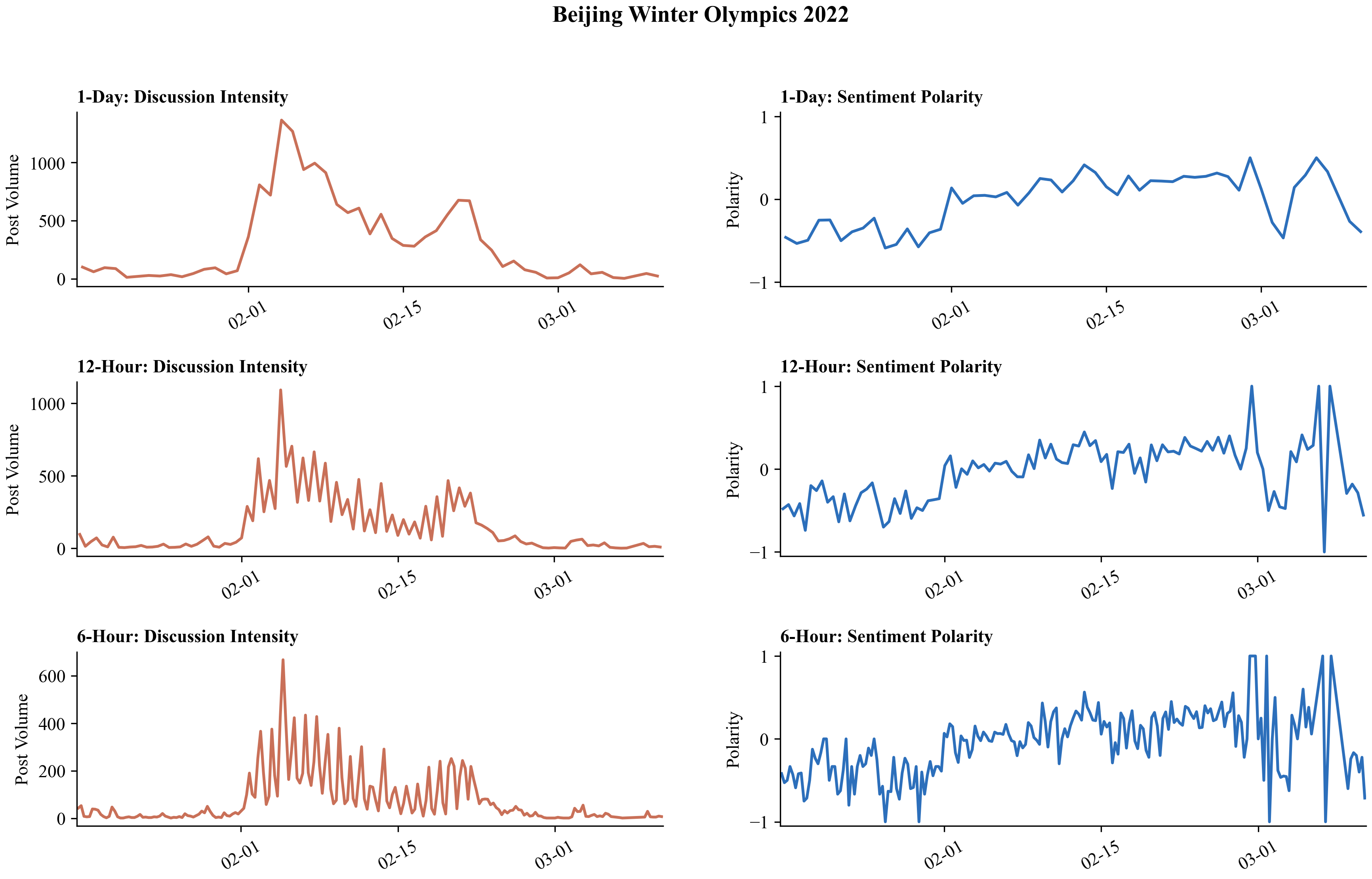}
\caption{Discussion Intensity and Sentiment Polarity at 1-Day, 12-Hour, and 6-Hour granularities for Beijing Winter Olympics 2022.}
\label{fig:appendix_beijing_olympics}
\end{figure}

\begin{figure}
\centering
\includegraphics[width=\textwidth]{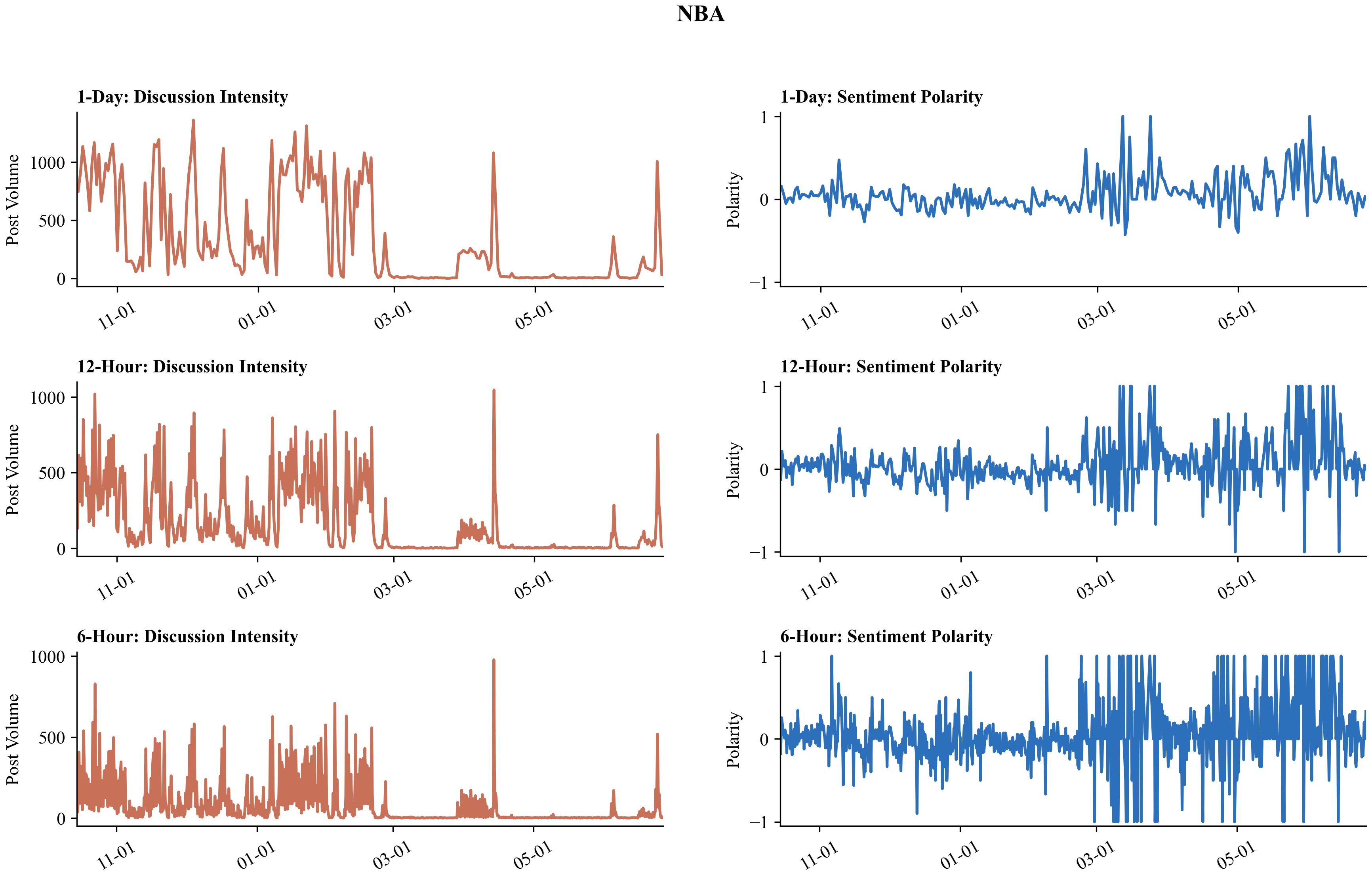}
\caption{Discussion Intensity and Sentiment Polarity at 1-Day, 12-Hour, and 6-Hour granularities for NBA.}
\label{fig:appendix_nba}
\end{figure}

\begin{figure}
\centering
\includegraphics[width=\textwidth]{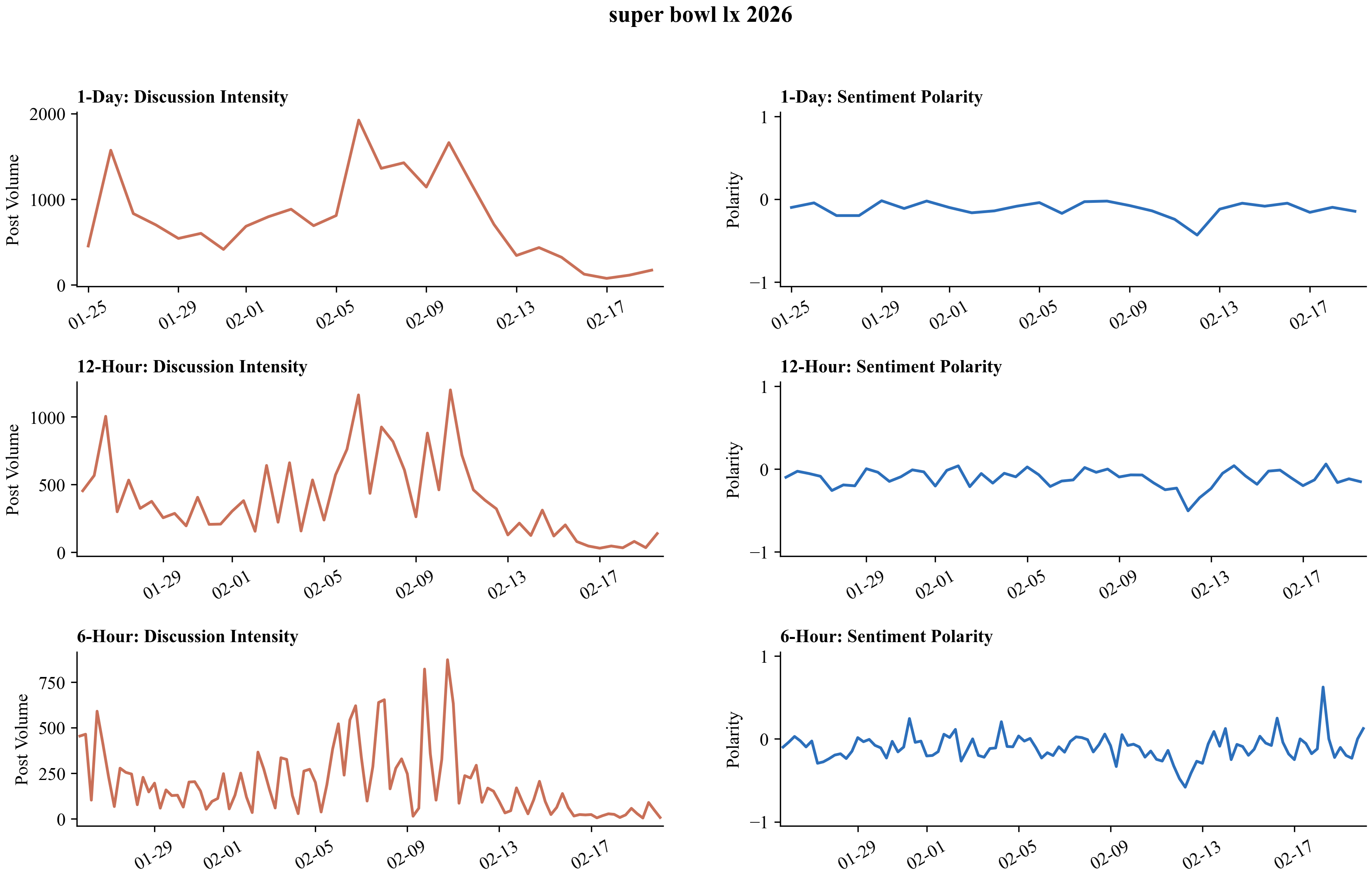}
\caption{Discussion Intensity and Sentiment Polarity at 1-Day, 12-Hour, and 6-Hour granularities for Super Bowl LX 2026.}
\label{fig:appendix_super_bowl}
\end{figure}

% 附录 H：伦理与隐私
%
\section{Ethics and Privacy}
\label{sec:appendix_ethics}

We accompany this section with a Datasheet \cite{gebru2021datasheets} and a partial Data Statement \cite{bender2018datastatements}, provided in Appendix~\ref{sec:appendix_datasheet}, which together document SURGE's motivation, composition, collection, preprocessing, intended uses, distribution, and maintenance.

\paragraph{Data Source Compliance.}
SURGE is constructed from social media posts on Twitter, Reddit, and Threads that were publicly accessible through each platform's documented public, search, or API interfaces available to the authors at acquisition time, in compliance with the then-current platform terms of service and API usage policies. Authentication tokens, where required by a platform's API, were used solely under their permitted use; posts that users had marked as private, restricted, or deleted are excluded. Rate limits and per-endpoint usage caps are observed throughout each acquisition campaign.

\paragraph{Privacy Protection.}
To protect user privacy, user identifiers are anonymized before release, and no personally identifiable information including names, profile URLs, contact details, or geolocation is included in the published dataset.
SURGE does not redistribute raw or sampled post text. Each bin's textual selection (top-3 main posts ranked by reply count plus up to 2 replies each, truncated to 1{,}500 characters when reconstructed) is released only as anonymized post IDs; users with platform access can reconstruct the text views locally via the released hydration script (Appendix~\ref{sec:appendix_ts_construction}). A small number of illustrative reconstructed examples appear in Appendix~\ref{sec:appendix_visualizations} for paper documentation only.
This release model substantially reduces redistribution of user-generated content while preserving benchmark reproducibility through pinned post-ID selections.

\paragraph{Removal Requests.}
Users whose anonymized content appears in the release may request removal by contacting the dataset maintainers through the channel listed in the dataset repository.
Confirmed requests are honored in subsequent dataset versions, and a changelog entry records the version in which the affected content is excluded.

\paragraph{Intended Use and Risk Considerations.}
SURGE covers events that include armed conflicts (e.g., Russia--Ukraine Conflict 2022, Israel--Hamas Conflict 2025, India--Pakistan Conflict), natural disasters, and political controversies, and the associated sentiment labels reflect the emotional tenor of public discourse around these topics. The intended use of this dataset is academic research on event dynamics, sentiment forecasting, and opinion diffusion modeling. Researchers should not use it for surveillance, individual targeting, harassment of communities, or any operational decision-making that targets identifiable users or groups.

\paragraph{Sentiment Labels Are Not Stance Labels.}
A general-purpose LLM produces post-level sentiment labels by reading expressed affect (positive, neutral, negative) without modelling stance toward a target, sarcasm, irony, or identity-targeted abuse. For conflict-heavy events these distinctions matter: a post critical of one party may read as ``negative'' while supporting another party's stance, and identity-targeted abuse is sometimes phrased in superficially positive language. We emphasize that the SURGE Sentiment Polarity series should be interpreted as polarity-of-expression aggregated over a bin rather than as a stance measurement, and downstream analyses on conflict-related events should be paired with stance-specific classifiers or domain-tuned validators before any policy-relevant claim is made. Aggregation across tens to hundreds of posts per bin attenuates per-post noise but does not correct systematic stance--sentiment confounds.

\paragraph{Demographic and Linguistic Coverage.}
The English-only scope (Section~\ref{sec:limitations}) limits demographic coverage to English-speaking discourse on Twitter, Reddit, and Threads in the 2022--2026 window. SURGE therefore only covers Anglophone publics and platform-specific user bases, and conclusions drawn from it should not be generalized to non-English communities or to platforms with different moderation regimes without explicit re-validation.

\paragraph{License and Release Artifacts.}
We separate the licensing of the artifacts we ourselves create from the status of the underlying user-generated content, which we cannot relicense. \emph{Author-created derivative metadata}---per-event time series at three granularities, anonymized per-bin post-ID selections, anonymized interaction edges, normalization statistics, and event metadata---are released under the Creative Commons Attribution 4.0 International (CC BY 4.0) license, with attribution required. \emph{Underlying post text} is not redistributed by SURGE; users who reconstruct text locally from the released anonymized post IDs via the hydration script are responsible for compliance with each platform's terms of service, including respecting upstream deletion and removal. \emph{Code} for the construction pipeline, baselines, the CMA reference baseline, and the hydration script is released under a permissive open-source license. Hosting, versioning, checksums, schema documentation, and example records are provided in the dataset repository, together with a changelog that tracks dataset versions and removal-request applications. Long-term maintenance is committed for a minimum window after the release; the maintenance plan, including the frequency of removal-request review and the policy for upstream-deletion propagation, is documented in the repository's data card.

% 附录 I：Datasheet
%
\section{Datasheet for SURGE}
\label{sec:appendix_datasheet}

We provide a Datasheet \cite{gebru2021datasheets} and partial Data Statement \cite{bender2018datastatements} for SURGE. This appendix is intended to be reproducible from the released artifacts and to support informed downstream use.

\subsection{Motivation}
\textbf{For what purpose was the dataset created?}
SURGE was created as a multi-event social media benchmark that pairs event-level sentiment time series with bin-aligned text and reply-and-repost interaction structure, in order to support event-driven sentiment forecasting research that is currently impeded by the absence of such a unified resource.

\textbf{Who created and who funded the dataset?}
Author affiliations and funding are listed in the camera-ready version of the manuscript and in the dataset repository.

\subsection{Composition}
\textbf{What do the instances represent?}
Each instance is one (event, bin) record at a given temporal granularity. A record contains: per-bin numerical targets (Discussion Intensity and Sentiment Polarity, in both raw and per-event z-score normalized form), a flat textual view, a structured textual view, and the bin-aligned reply/repost edges between sampled posts.

\textbf{How many instances are there in total?}
The release contains 67 events covering 817{,}442 posts, organized into 67 events at 6-hour granularity, 64 events at 12-hour granularity, and 55 events at 1-day granularity.

\textbf{Does the dataset contain all possible instances or is it a sample?}
Sampled. From a raw collection of 1{,}256{,}816 posts and 93 candidate events, SURGE retains 67 events and 817{,}442 posts after the deduplication and quality filtering steps documented in Appendix~\ref{sec:appendix_data_preprocessing} together with the active-period refinement step documented in Appendix~\ref{sec:appendix_ts_construction}. Per bin, a fixed selection of posts (the top-3 main posts ranked by reply count and up to 2 replies each, truncated to 1{,}500 characters when reconstructed) is released as anonymized post IDs only; the corresponding text content is not redistributed and is instead reconstructed locally via the released hydration script. A small number of illustrative reconstructed examples appear in Appendix~\ref{sec:appendix_visualizations} for paper documentation only. The full reply/repost edge list is released separately as anonymized edges for users who require the unsampled graph.

\textbf{Is there a label or target associated with each instance?}
Yes, the two target variables defined in Equation~\ref{eq:targets}.

\textbf{Is any information missing?}
Bins that contain no posts are kept as missing (NaN) in the released CSV files. The benchmark pipeline imputes them via forward fill within each split independently at load time, so no information crosses split boundaries (Appendix~\ref{sec:appendix_ts_construction}).

\textbf{Are relationships between instances explicit?}
Yes. Reply and repost edges are released as an explicit edge list per event, with each edge annotated by source post, target post, edge type, and timestamp.

\textbf{Are there recommended data splits?}
Yes. Each event is split chronologically into 70\% training, 10\% validation, and 20\% test segments. For cross-category generalization the recommended protocol is leave-one-category-out over the five event categories.

\textbf{Are there errors, sources of noise, or redundancies?}
Yes. (i) Sentiment labels are produced by a general-purpose LLM (Qwen3-32B) and may inherit its biases on stance, sarcasm, and identity-targeted abuse (Appendix~\ref{sec:appendix_ethics}). (ii) Forward-fill imputation within each split can extend trends across silent bins inside that split.

\textbf{Does the dataset contain confidential or sensitive content?}
Yes, in the sense that posts about armed conflicts, natural disasters, and political controversies appear. All posts were publicly accessible through each platform's documented interfaces at acquisition time, and posts that users had subsequently marked as private, restricted, or deleted are excluded.

\subsection{Collection Process}
\textbf{How was the data acquired?}
Through each platform's documented public, search, or API interfaces available to the authors between 2022 and 2026, in independent acquisition campaigns each targeting an event or topical cluster. See Appendix~\ref{sec:appendix_data_collection} for the campaign-level breakdown.

\textbf{Was any consent obtained?}
No individual consent was obtained, in line with standard practice for publicly accessible social media research. We protect users through anonymization, content sampling, and a removal-request mechanism (Appendix~\ref{sec:appendix_ethics}).

\subsection{Preprocessing, Cleaning, and Labelling}
The full preprocessing pipeline is documented in Appendix~\ref{sec:appendix_data_preprocessing}.

\subsection{Uses}
\textbf{Has the dataset been used for any tasks already?}
The benchmark experiments in Section~\ref{sec:exp_setup} cover numerical-only forecasting, text-augmented forecasting, structure-aware evaluation, and leave-one-category-out cross-category generalization across ten numerical TSF baselines, three multimodal TSF baselines, and the CMA structure-aware reference baseline.

\textbf{Are there tasks for which the dataset should not be used?}
Surveillance, individual targeting, harassment of communities, and any operational decision-making that targets identifiable users or groups (Appendix~\ref{sec:appendix_ethics}).

\subsection{Distribution}
\textbf{How will it be distributed?}
Through the dataset repository linked from the camera-ready manuscript, hosted on a stable academic-data platform with versioning and checksums.

\textbf{When will it be released and under what license?}
At the time of camera-ready submission, with author-created derivative metadata under CC BY 4.0 and code under a permissive open-source license. Underlying user-generated text remains the property of its original authors and is governed by each platform's terms of service (Appendix~\ref{sec:appendix_ethics}).

\textbf{What exactly is released, per platform?}
Table~\ref{tab:release_artifacts} enumerates the artifact components and their per-platform release status, license footing, and refresh/takedown behaviour. The same table also makes explicit what is \emph{not} redistributed: raw or sampled post text in any form, raw user identifiers, profile metadata, and any geolocation. Text views can be reconstructed locally via the released hydration script applied to the released anonymized post IDs.

\begin{table}[h]
\centering
\caption{Release artifacts and their per-platform status. SURGE does not redistribute raw or sampled post text; text views are reconstructed locally by users with platform access via the released hydration script applied to the released anonymized post IDs. ``Refreshable'' indicates components regenerated at each version bump to reflect upstream deletions.}
\label{tab:release_artifacts}
\small
\begin{tabular}{l c c c c}
\toprule
\textbf{Artifact component} & \textbf{Twitter} & \textbf{Reddit} & \textbf{Threads} & \textbf{License footing} \\
\midrule
Per-bin numerical targets ($c_t$, $\bar{s}_t$) & \checkmark & \checkmark & \checkmark & CC BY 4.0 (derivative) \\
Per-event normalization statistics & \checkmark & \checkmark & \checkmark & CC BY 4.0 (derivative) \\
Sampled per-bin post-ID selections & \checkmark & \checkmark & \checkmark & CC BY 4.0 \\
Reply / repost edge list (per event), anonymized & \checkmark & \checkmark & \checkmark & CC BY 4.0 \\
Anonymized post identifiers & \checkmark & \checkmark & \checkmark & CC BY 4.0 \\
Hydration script for text-view reconstruction & \multicolumn{3}{c}{Permissive open-source} & --- \\
Raw or sampled post text & --- & --- & --- & not redistributed \\
User identifiers / profile metadata & --- & --- & --- & not released \\
Geolocation / device metadata & --- & --- & --- & not released \\
Construction and benchmark code & \multicolumn{3}{c}{Permissive open-source} & --- \\
\bottomrule
\end{tabular}
\end{table}

\subsection{Annotation Validation}
The released sentiment series is produced by a general-purpose LLM (Qwen3-32B) and is treated as a reproducible LLM-derived signal rather than a gold-standard human label (Section~\ref{sec:limitations} L3). A stratified human verification study on $3{,}000$ posts ($200$ per category-class cell across the five event categories and three sentiment classes) is reported in Appendix~\ref{sec:appendix_sentiment}, including per-cell agreement, per-class F1, and the per-stratum systematic-bias estimate $|\mu_{c,k}| \leq 0.05$. Downstream users who run conflict-related analyses should still pair SURGE's bin-level Sentiment Polarity with stance-specific or domain-tuned validators rather than treat it as a direct measurement of public stance.

\subsection{Maintenance, Deletion, and Version Semantics}
\textbf{Who will maintain the dataset?}
The dataset maintainers listed in the repository.

\textbf{Deletion and rehydration semantics.}
The released sampled bin-level text views and the per-event reply/repost edge lists are tied to upstream platform state. At each version release, the maintainers re-query the upstream identifiers of all included posts. Posts that have been deleted, made private, restricted, or removed by the user or by the platform are removed from the next version's released text views, and any reply/repost edges that point to such posts are dropped from the released graph. Dataset users do not need to perform their own rehydration: stored text in each released version reflects the upstream state at the time of that version's snapshot, so deletions propagate at version-bump granularity rather than in real time. Users who require finer-grained synchronization with upstream deletion should re-query upstream platforms directly using the released anonymized post identifiers.

\textbf{Versioning.}
Each release is tagged with a SemVer-style identifier. Patch versions correct labeling or metadata bugs without changing the set of included events or the post-sampling design. Minor versions add additional events or additional textual fields without breaking the schema. Major versions change the schema or the sampling design (e.g., changing $K_{\text{post}}$ or $K_{\text{reply}}$). Each version ships a changelog that lists: (i) the cumulative number of removal requests applied since release, (ii) the upstream-deletion delta relative to the previous version, (iii) any schema or sampling-design changes, and (iv) the version of the construction pipeline used. Reproducing a previously published benchmark result requires pinning to the corresponding dataset version identifier, which we recommend authors cite explicitly.

\textbf{Removal request channel.}
Users whose anonymized content appears in the release may request removal by contacting the maintainers via the channel listed in the repository. Confirmed requests are honored in the next release, with a changelog entry recording the version in which the affected content is excluded.

\end{document}